\documentclass[12pt]{article}
\input epsf.sty
\usepackage{mathrsfs}
\usepackage{amsmath}
\usepackage{mathrsfs}
\usepackage{amssymb}
\usepackage{color}
\usepackage{psfrag}
\usepackage{graphicx}
\usepackage{subfigure}
\usepackage{rotating}

\newcommand{\bea}{\begin{eqnarray}}
\newcommand{\eea}{\end{eqnarray}}
\newcommand{\be}{\begin{equation}}
\newcommand{\ee}{\end{equation}}
\newcommand{\vs}[1]{\vspace{#1 mm}}

\newcommand{\dsl}{\pa \kern-0.5em /}

\newcommand{\pa}{\partial}

\newcommand{\nn}{\nonumber\\}

\begin{document}
\topmargin 0pt
\oddsidemargin 0mm




\vspace{2mm}

\begin{center}

{\Large \bf {Wilson loops in noncommutative Yang-Mills theory\\
using gauge/gravity duality}}

\vs{10}

{Somdeb Chakraborty\footnote{E-mail: somdeb.chakraborty@saha.ac.in},
Najmul Haque\footnote{E-mail: najmul.haque@saha.ac.in} and 
Shibaji Roy\footnote{E-mail: shibaji.roy@saha.ac.in}}

 \vspace{4mm}

{\em

 Saha Institute of Nuclear Physics,
 1/AF Bidhannagar, Calcutta-700 064, India\\}

\end{center}

\vs{10}

\begin{abstract}
By using the gauge/gravity duality and the Maldacena prescription we compute
the expectation values of the Wilson loops in hot, noncommutative 
Yang-Mills (NCYM)
theory in (3+1) dimensions. We consider both the time-like and the light-like
Wilson loops. The gravity dual background is given by a particular decoupling
limit of non-extremal (D1,D3) bound state of type IIB string theory. We obtain 
the velocity
dependent quark-antiquark potential and numerically study how the dipole
length and the potential change with velocity (for $0<v<1$, i.e., the Wilson 
loop is time-like) of the dipole as well as noncommutativity. We discuss and 
compare the results with the known commutative results. We also obtain 
an analytic expression for the screening length when the rapidity is large and 
the noncommutativity parameter is small with the product remaining small. When
$v \to 1$, the time-like Wilson loop becomes light-like and in that case we
obtain the form of the jet quenching parameter for the strongly coupled
noncommutative Yang-Mills plasma which matches with our earlier results
obtained using different approach.     
\end{abstract}

\newpage

\section{Introduction}

One of the remarkable features of AdS/CFT correspondence 
\cite{Maldacena:1997re,Witten:1998qj,Gubser:1998bc} and its generalizations 
\cite{Aharony:1999ti} is that it gives us access to the
non-perturbative regimes of large $N$ gauge theories simply from
the low energy, weakly coupled string theory in certain backgrounds.
So, for example, the expectation values of Wilson loops, which are 
non-perturbative 
objects in gauge theories, can be computed using AdS/CFT correspondence as 
has been prescribed in
\cite{Maldacena:1998im,Rey:1998ik,Rey:1998bq,Brandhuber:1998bs}.
In strongly coupled gauge theories 
of interacting quark-gluon plasma, Wilson loops can be related to 
various measurable quantities in heavy ion experiments in RHIC or in LHC.
For example, the expectation value of a special time-like Wilson loop can be
related to the static quark-antiquark potential \cite{Wilson:1974sk} in 
a moving quark-gluon
plasma. On the other hand, the expectation value of a particular light-like
Wilson loop can be related, among other things, to the radiative energy loss
of a parton or the jet quenching parameter \cite{Kovner:2003zj}. 

The velocity dependent quark-antiquark potential of a dipole moving with an 
arbitrary
velocity through the hot quark-gluon plasma including the screening length 
\cite{Liu:2006nn,Caceres:2006ta,Chernicoff:2006hi,Avramis:2006em} as well 
as the jet quenching parameter  
\cite{Liu:2006ug,Liu:2006he}\footnote{Also see \cite{CasalderreySolana:2011us}
for a recent review.} have been calculated when the
plasma is described by $D=4$, ${\cal N} = 4$, SU($N$) Yang-Mills theory 
using AdS/CFT correspondence\footnote{Jet quenching parameter in various other 
theories have been obtained in \cite{Buchel:2006bv}. Also the drag force 
on a moving quark have been calculated in \cite{Herzog:2006gh}.}.  
In this
paper we calculate the velocity dependent quark-antiquark potential, 
the screening length
and the jet quenching parameter when the plasma is described by $D=4$, thermal,
noncommutative Yang-Mills (NCYM)\footnote{   
The space-time noncommutativity is an old idea introduced first by Heisenberg
and Pauli \cite{Heisenberg:1929xj} in order to avoid infinities in 
quantum field theory before 
the renormalization was successful. It was Snyder \cite{Snyder:1946qz} and 
then Connes \cite{Connes:1994} who took the idea seriously. Connes along with
Chamseddine \cite{Chamseddine:1991qh} even introduced noncommutative geometry 
as a generalization to Riemannian geometry and obtained gauge theory
as a companion to general relativity giving rise to a true geometric 
unification. In this framework parameters of the standard model appear as 
geometric invariants. Even though a consistent gauge theory can be formulated
in noncommutative space-time, so far, there is no evidence for its existence 
in nature in low energy. However one can not rule out the possibility that
its effect could be detected in some future experiments. The experimental 
lower bound on the noncommutativity
scale reported in the literature \cite{Carroll:2001ws} 
usually gives a very small effect and is hard
to detect. So, it is desirable to look for its effect in as many different
cases as possible. High energy heavy ion collision is one such possible case
and it may be worth while to look whether it can provide a better window for 
the effect of space-time noncommutativity to be observed.       
One might wonder how would the space-time noncommutativity appear in the heavy
ion collision in the first place? It is known that one of the mechanisms 
for the appearance of spatial
noncommutativity is the presence of an intense magnetic field in the 
background. It has been shown in both analytic calculations 
\cite{Kharzeev:2007jp} and numerical simulations \cite{Skokov:2009qp} 
that such an intense magnetic field is indeed possible in 
the heavy ion collision in RHIC (or in LHC). So, it may be
quite relevant to consider such a possibility in the present context.}
theory at large $N$ using gauge/gravity duality. 
NCYM theory arises quite naturally in string theory
\cite{Seiberg:1999vs, Maldacena:1999mh, Hashimoto:1999ut} 
and M-theory \cite{Connes:1997cr} and it is of
interest to see how noncommutativity affects the known velocity dependent
quark-antiquark potential, screening length as well as the jet quenching
parameter of the ordinary super YM theory.   

The gravity background in this case is given by a particular 
decoupling limit \cite{Maldacena:1999mh} of non-extremal (D1, D3) 
bound state system of type IIB 
string theory. (D1, D3) bound state \cite{Breckenridge:1996tt, Cai:2000hn} 
contains a non-zero $B$-field and it 
becomes asymptotically very large in the decoupling limit and is the source
of space-space noncommutativity \cite{Seiberg:1999vs}. We use the string 
probe in this background
and extremize the Nambu-Goto string world-sheet action in a particular 
static gauge and in turn obtain the expectation value of the Wilson loop, 
where the loop is the boundary of the above minimal area. We consider both
the time-like and the light-like Wilson loops. From the time-like Wilson
loop, we calculate the velocity dependent quark-antiquark potential, when
the background or the plasma is moving with a velocity $v$ (for $0<v<1$) and
the end-points of the fundamental string act as a heavy quark-antiquark pair
or a dipole in the boundary gauge theory. We will take the background (or the 
plasma) to be moving along one of the brane directions (which is taken to
be a commutative direction) relative to the dipole which lies along one
of the noncommutative directions. There are other possibilities, but this is
the simplest case where the noncommutative effect shows up. The 
quark-antiquark potential obtained
from the minimal area of the world-sheet can only be given numerically.
We first plot the dipole length as a function of certain constants of motion 
and then using this we plot the potential as a function of the dipole length
at different velocities and noncommutativity parameters. The results are 
compared 
with the known commutative case \cite{Liu:2006he}. We also give an analytic 
expression for
the screening length when the rapidity is large and the noncommutativity 
parameter
is small with the product also remaining small. Finally, for the completeness,
we consider the case where the velocity
$v \to 1$ and the time-like Wilson loop becomes light-like. In this case
we recover the form of the jet quenching parameter as obtained before 
\cite{Chakraborty:2011gn} using a different approach. 

This paper is organized as follows. In section 2, we compute the time-like
Wilson loop and give our numerical results for the dipole length and the 
quark-antiquark potential. We also obtain the analytic form of the screening 
length in some spacial case. In section 3, we take the velocity to unity and
obtain the light-like Wilson loop. From this we obtain the form of the
jet quenching parameter in this theory. Finally, we conclude in section 4.

\section{Time-like Wilson Loop and $Q$-$\bar Q$ 
potential}  

In this section we will compute the time-like Wilson loop for the hot
noncommutative Yang-Mills theory in (3+1)-dimensions from gauge/gravity
duality. The gravity dual of noncommutative Yang-Mills theory is given by a
particular decoupling limit \cite{Maldacena:1999mh, Hashimoto:1999ut} of 
the non-extremal (D1, D3) bound state of type
IIB string theory. We use the fundamental string as a probe in this 
background and compute the Nambu-Goto string world-sheet action. This action 
is then extremized to relate it to the expectation value of the time-like 
Wilson loop \cite{Liu:2006he}. We will study numerically both the 
separation length of the 
quark-antiquark and the potential when we vary the velocity and the 
noncommutativity parameter. We will also give an analytic expression of the
screening length in some special limit.

The non-extremal (D1, D3) bound state solution of type IIB string theory
is given by the following metric (given in the string frame), the dilaton, the
NSNS $B$-field and the RR form-fields \cite{Cai:2000hn},
\bea\label{d1d3}
ds^2 &=& H^{-\frac{1}{2}}\left(-f(dt)^2 + (dx^1)^2 + \frac{H}{F}\left((dx^2)^2
+ (dx^3)^2\right)\right) + H^{\frac{1}{2}}\left(\frac{dr^2}{f} + r^2
d\Omega_5^2\right)\nn
e^{2\phi} &=& g_s^2 \frac{H}{F},\qquad B_{23} = \frac{\tan\alpha}{F}\nn
A_{01} &=& \frac{1}{g_s} (H^{-1}-1)\sin\alpha\coth\varphi,\qquad
A_{0123} = \frac{1}{g_s}\frac{(1-H)}{F}\cos\alpha \coth\varphi + {\rm T.\,T.}
\eea
where the various functions appearing above are,
\be\label{fns}
f = 1 - \frac{r_0^4}{r^4},\qquad H = 1 + \frac{r_0^4 \sinh^2 \varphi}{r^4},
\qquad F = 1 + \frac{r_0^4 \cos^2\alpha \sinh^2\varphi}{r^4}
\ee
Here D3-branes lie along $x^1,\,x^2,\,x^3$ and D1-branes lie along $x^1$. 
The angle $\alpha$ measures the relative number of D1 and D3 
branes by the relation $\cos\alpha = N/\sqrt{N^2+M^2}$, where $N$ is 
the number of D3-branes and $M$ is the number of D1-branes per unit 
codimension two surface transverse to D1-branes \cite{Lu:1999uv}. Also in 
the above $\varphi$
is the boost parameter and $r_0$ is the radius of the horizon of the
non-extremal (D1, D3) bound state solution. $\phi$ is the dilaton and 
$g_s$ is the string coupling constant. $A_{01}$ and $A_{0123}$ are the RR
form-fields corresponding to D1-brane and D3-brane respectively. T.T.
denotes a term, involving transverse part of the brane to make the 
field-strength self-dual, whose explicit form is not required for our
discussion. $B_{23}$ is the NSNS form responsible for the appearance of 
noncommutativity in the decoupling limit.

The NCYM decoupling limit is a low energy limit for which we
zoom into the region \cite{Maldacena:1999mh},
\be\label{ncymlimit}
r_0 < r \sim r_0 \sqrt{\sinh\varphi\cos\alpha} \ll r_0 \sqrt{\sinh\varphi}
\ee
The above limit implies that $\varphi$ is a large parameter and the angle
$\alpha$ is close to $\pi/2$. In this limit we get,
\be\label{approx}
H \approx \frac{r_0^4 \sinh^2\varphi}{r^4}, \qquad \frac{H}{F} \approx 
\frac{1}{\cos^2\alpha(1+a^4r^4)} \equiv \frac{h}{\cos^2\alpha}
\ee
where we have defined
\be\label{def}
h \equiv \frac{1}{1+a^4r^4}, \qquad {\rm with}, \quad a^4 \equiv 
\frac{1}{r_0^4\sinh^2\varphi \cos^2\alpha}
\ee
From \eqref{d1d3} we notice that since the asymptotic value of $B$-field
is $\tan \alpha$ and $\alpha \to \pi/2$ in the decoupling limit, the
$B$-field becomes very large. The non-vanishing component of the $B$-field
is $B_{23}$ which gives rise to a magnetic field in the D3-brane world-volume
and is responsible for making $x^2$ and $x^3$ directions noncommutative 
\cite{Ardalan:1999av}.
Using \eqref{approx}, we rewrite the metric in \eqref{d1d3} as,
\be\label{d1d3one}
ds^2 = \frac{r^2}{r_0^2\sinh\varphi}\left(-f dt^2 + (dx^1)^2 +
  h \left[(dx^2)^2+(dx^3)^2\right]\right)
+ \frac{r_0^2\sinh\varphi}{r^2}\left(\frac{dr^2}{f} 
+ r^2 d\Omega_5^2\right)
\ee
Here the function $h$ is as defined in \eqref{def} and also in writing
\eqref{d1d3one} we have rescaled the coordinates as, $x^{2,\,3} \to \cos\alpha
x^{2,\,3}$.
This metric along with the other field configurations given in \eqref{d1d3}
in the NCYM decoupling limit is the gravity dual of four dimensional thermal 
NCYM theory. We use fundamental open string as a probe and consider its
dynamics in this background. Let the line joining the end points of the string
or the dipole lie along $x^2$ one of the noncommutative directions
and move along $x^1$ with a velocity $v$ where $0<v<1$ \footnote{There 
are various
other possibilities one can consider, for example, the dipole lies along the
commutative direction $x^1$ and moves along one of the noncommutative 
directions $x^2$ (say) or the dipole lies along one of the noncommutative
directions $x^2$ and moves along the other noncommutative direction $x^3$.
Dipole can even have an arbitrary orientation with respect to its motion
and the motion can also be in arbitrary direction in the mixed 
commutative-noncommutative boundary.  
Here we consider only the simplest case to see the noncommutative effect.}. 
To simplify
the calculation we can go to the rest frame $(t',\,x^1\,')$ of the dipole by 
boosting the coordinate as,
\bea\label{boost}
dt &=& \cosh \eta dt' - \sinh \eta dx^1\, '\nn
dx^1 &=& -\sinh \eta dt' + \cosh \eta dx^1\, '
\eea
where the rapidity $\eta$ is related to $v$ as $\tanh \eta = v$. So, in this
frame the dipole is static and the background (or the plasma) is moving with
a velocity $v$ in the negative $x^1$-direction. Note that the rectangular Wilson
loop lies along $t'$ and $x^2$ directions and we denote the lengths along
those directions as ${\cal T}$ and $L$ respectively. We also assume ${\cal
  T}\gg L$ such that the string world-sheet is time translation invariant.
Using \eqref{boost} in the metric \eqref{d1d3one} we get,
\bea\label{d1d3two}
ds^2 &=& -A(r)dt^2 - 2B(r) dt dx^1 + C(r) (dx^1)^2 + \frac{r^2
  h}{r_0^2\sinh\varphi} \left[(dx^2)^2+(dx^3)^2\right]\nn
& & \qquad\qquad\qquad + \frac{r_0^2 \sinh\varphi}{r^2}\frac{dr^2}{f} + r_0^2
\sinh\varphi d\Omega_5^2\nn
&=& G_{\mu \nu}dx^{\mu} dx^{\nu}
\eea
where
\bea\label{abc}
A(r) &=& \frac{r^2}{r_0^2\sinh\varphi}\left(1 - \frac{r_0^4 \cosh^2\eta}{r^4}
\right)\nn      
B(r) &=& \frac{r_0^2 \sinh\eta \cosh\eta}{r^2\sinh\varphi}\nn
C(r) &=& \frac{r^2}{r_0^2\sinh\varphi}\left(1 + \frac{r_0^4 \sinh^2\eta}{r^4}
\right)
\eea 
Note that since we will be using the `primed' coordinates from now on, we have
dropped the prime for simplicity. We will evaluate the Nambu-Goto action of
the string world-sheet in this background. The Nambu-Goto action is given as,
\be\label{ng}
S = \frac{1}{2\pi\alpha'}\int d\sigma d\tau \sqrt{-{\rm det}\,g_{\alpha\beta}}
\ee
where $g_{\alpha\beta}$ is the induced metric on the world-sheet given as
\be\label{induced}
g_{\alpha\beta} = G_{\mu\nu} \frac{\partial x^\mu}{\partial \xi^\alpha}
\frac{\partial x^\nu}{\partial \xi ^\beta}
\ee     
with $G_{\mu\nu}$ is as given in \eqref{d1d3two}. In the above $\xi^{\alpha}$
are the world-sheet coordinates $\tau$ and $\sigma$ for $\alpha=0$ and $\alpha
=1$ respectively. We choose the static gauge for evaluating \eqref{ng} as,
$\tau = t$, $\sigma = x^2$, where $-L/2 \leq x^2 \leq L/2$ and $r=r(\sigma)$,
$x^1(\sigma) = x^3(\sigma) = $ constant. $r(\sigma)$ is the string embedding
we want to determine by extremizing the Nambu-Goto action, with the boundary 
condition $r(\pm L/2) = r_0\Lambda$ (here $\Lambda$ is a parameter). Using
these in the action \eqref{d1d3two}, we get,
\be\label{ngone}
S = \frac{{\cal T}}{2\pi\alpha'}\int_{-L/2}^{L/2} d\sigma \left[A(r)\left(
\frac{r^2 h}{r_0^2\sinh\varphi} + \frac{r_0^2
  \sinh\varphi}{r^2}\frac{\left(\partial_\sigma
    r\right)^2}{f}\right)\right]^{\frac{1}{2}} 
\ee
with $A(r)$ as given in \eqref{abc}. Introducing dimensionless quantities
$y = r/r_0$, $\tilde \sigma = \sigma/(r_0\sinh\varphi)$ and $\ell =
L/(r_0\sinh\varphi)$, the Nambu-Goto action \eqref{ngone} can be rewritten as,
\be\label{ngtwo}
S = \frac{{\cal T} r_0}{\pi\alpha'}\int_0^{\ell/2} d\sigma {\cal L} =
{\cal T} T \sqrt{\hat \lambda} \int_0^{\ell/2} d\sigma {\cal L}
\ee
where
\be\label{lagrangian}
{\cal L} = \sqrt{\left(y^4-\cosh^2\eta\right)\left(\frac{1}{1+a^4r_0^4y^4} +
\frac{y'^2}{y^4-1}\right)}
\ee
Note that we have removed `tilde' from $\tilde \sigma$ while writing
\eqref{ngtwo} as it is an integration variable. Also $y'$ denotes
$dy/d\sigma$ and we have used the fact that $y$ is an even function of $\sigma$
by symmetry. In writing the second expression in \eqref{ngtwo} we have made use
of the standard gauge/gravity relation 
\cite{Maldacena:1999mh,Hashimoto:1999ut},
\bea\label{gaugegravity}
& & T = \frac{1}{\pi r_0 \cosh\varphi} \approx 
\frac{1}{\pi r_0 \sinh\varphi} , \qquad r_0^4 \sinh^2\varphi = 
2 \hat g_{\rm
  YM}^2 N \alpha'^2 = \hat \lambda \alpha'^2\nn 
& & {\rm and} \quad
a^4 r_0^4 \equiv \frac{1}{\sinh^2\varphi \cos^2\alpha} = \pi^4 \hat \lambda T^4
\theta^2 
\eea
The first relation in \eqref{gaugegravity} is obtained from calculating the 
Hawking temperature
of the non-extremal (D1, D3) brane given by the metric in \eqref{d1d3}
(note that in the decoupling limit when $\varphi$ is large 
$\cosh\varphi \approx \sinh\varphi$) and this
is the temperature of the NCYM theory by gauge/gravity duality. The second
relation is obtained from the D3-brane charge where $N$ is the number of
D3-branes and in NCYM theory this is the rank of the gauge group. $\hat g_{\rm
  YM}$ is the NCYM coupling and $2\hat g_{\rm YM}^2 N = \hat \lambda$
\footnote{We have used the convention adopted in 
\cite{Liu:2006ug, Liu:2006he}.} 
is the
't Hooft coupling of the NCYM theory. The NCYM 't Hooft coupling is related to
the ordinary 't Hooft coupling by the relation $\lambda = (\alpha'/\theta)\hat
\lambda$, where $\theta$ is the noncommutativity parameter defined by
$[x^2,\,x^3] = i\theta$ \cite{Maldacena:1999mh}. Here $\theta$ is a finite 
parameter and in the
decoupling limit as $\alpha' \to 0$, $\hat \lambda$ remains finite. 
The third relation is obtained using the first two and also using $\cos\alpha
= \alpha'/\theta$ in the decoupling limit. Note that in the decouping limit 
as $\alpha' \to 0$, $\alpha \to \pi/2$ as mentioned earlier. We will
compute $y(\sigma)$ by extremizing the action \eqref{ngtwo}. 

Now since the Lagrangian density in \eqref{ngtwo} does not explicitly depend
on $\sigma$, we have the following constant of motion,
\be\label{const}
{\cal H} = {\cal L} - y'\frac{\partial {\cal L}}{\partial y'}
= \frac{y^4 - \cosh^2\eta}{(1+ a^4r_0^4y^4)\sqrt{(y^4 - \cosh^2\eta) \left(
\frac{1}{1+a^4r_0^4y^4} + \frac{y'^2}{y^4 -1}\right)}} = q = {\rm const.}
\ee
As in the commutative theory \cite{Liu:2006he} we will consider two cases: 
(a) $\sqrt{\cosh\eta}
< \Lambda$ and then take $\Lambda \to \infty$. The rapidity in this case
remains finite, the Wilson loop is time-like and the action is real. We will
compute the quark-antiquark potential in this case and also give an expression
of screening length in some special case. (b) $\sqrt{\cosh\eta} > \Lambda$ and
then take $\eta \to \infty$, keeping $\Lambda$ finite. The Wilson loop in this
case is light-like and the action is imaginary. We will take $\Lambda \to
\infty$ in the end and obtain the expression of the jet quenching parameter
for the noncommutative hot Yang-Mills plasma.

We will consider case (a) in this section and case (b) in the next section.
When $\sqrt{\cosh\eta} < \Lambda$, the action would be real and from
\eqref{const} $y'$ can be solved as,
\be\label{yprime}
y' = \frac{\sqrt{1-a^4r_0^4q^2}}{q(1+a^4r_0^4y^4)}\sqrt{(y^4-1)\left(y^4 -
    y_c^4\right)}
\ee
where $y_c^4 = (\cosh^2\eta + q^2)/(1 - a^4r_0^4q^2) > 1$ denotes the larger
turning point where $y'$ vanishes. Integrating \eqref{yprime} we obtain,
\be\label{ellq}
2\int_0^{\ell/2} d\sigma = \ell(q) = \frac{2 q}{\sqrt{1-a^4r_0^4q^2}} 
\int_{y_c}^{\Lambda} \frac{1 + a^4 r_0^4 y^4}{\sqrt{(y^4-1)(y^4-y_c^4)}}
dy \ee
We remark that if we naively take $\Lambda$, where the boundary theory is
supposed to live to $\infty$, the above integral diverges. Here $\ell$ is
related to the dipole length $L$ by $\ell = L/(r_0\sinh\varphi) = \pi L T$
and so the divergence in $\ell(q)$ is physically meaningless. Note that
$\ell(q)$ in the commutative theory is indeed finite as can be seen 
from \eqref{ellq} by putting
$a^2r_0^2 \sim \theta$, which is a measure of noncommutativity, to
zero. However, for the noncommutative case $\ell(q)$ is divergent if we 
take $\Lambda \to \infty$. The reason is that the noncommutative gauge 
theory does
not live at $\Lambda = \infty$, but at some finite value of $\Lambda$, whose
exact value is not known. This is implicit in \cite{Maldacena:1999mh}.
It has been shown there that in a noncommutative theory it is not possible to 
fix the position of 
the string end point at infinity as a small perturbation would change it 
violently. This has also been explicitly mentioned in \cite{Dhar:2000nj}.

In the context of a Wilson loop calculation,
it has been noticed before \cite{Alishahiha:1999ci} that the string end 
points for a static string 
can not be fixed at a finite length at $\Lambda \to \infty$ in a
noncommutative theory. Therefore,
the dipole length $L$ indeed diverges. The reason for this divergence has
been argued to be the non-local interaction between the $Q$-$\bar Q$ pair in a
magnetic field \cite{Bigatti:1999iz}. To
be precise, the interaction point in terms of the center of mass coordinate
gets shifted by a momentum dependent term. Thus if the only non-zero component
of the $B$-field is $B_{23}$, as in our case, then by placing the dipole
along $x^2$, it automatically gets a momentum along $x^3$. So, if we keep the
dipole static along $x^3$, the length will diverge at infinity. To compensate
the momentum along $x^3$, the dipole must move along $x^3$ with a
particular velocity \cite{Haque:2009hz}. In that case, the end points of the 
string can be fixed at a finite length on the boundary at infinity and thus 
the divergence in the dipole length gets removed.             

In the following, we, however, remove the divergence in the dipole length in
a different way. We have noticed in \eqref{ellq}, the integral representation
of the dipole length, that by explicit evaluation the integral diverges
(note that the dipole in our case does not move along $x^3$) as we take
$\Lambda \to \infty$. But, the integral can be regularized to give a finite
result if we can explicitly extract the divergent term (the part that goes to
infinity as $\Lambda \to \infty$) of the integral and
remove it\footnote{We have checked explicitly, by doing the Wilson loop
  calculation in the zero temperature and zero rapidity ($\eta \to 0$, implying
  that the dipole is not moving along $x^1$) case and comparing it with
  eq.(45) of \cite{Alishahiha:1999ci} that  
  the two ways of regularizing (namely, by physically moving the dipole along
$x^3$ with a particular velocity \cite{Haque:2009hz} or by removing the 
divergent part of the
integral of $\ell(q)$) the dipole length in the noncommutative theory give 
identical results when the turning point ($y_c$) is large. However, for
small turning point dipole lengths obtained by these two methods differ by
some finite term. The same is
true for the calculation of $Q$-$\bar Q$ potential. We would like to thank the
anonymous referee for raising this issue.}. 
After regularization, the noncommutative theory may be thought of 
as living at
$\Lambda = \infty$. So, after regularization we will take $\Lambda \to 
\infty$. It is not difficult to obatin the divergent term from the
integral \eqref{ellq} when $\Lambda \to \infty$ and by inspection it can be
seen to have the form $2qa^4r_0^4\Lambda/\sqrt{1-a^4r_0^4q^2}$.
Subtracting the
divergent part the finite $\ell(q)$ can be written in the integral form as,
\be\label{regellq}
\ell(q) = \frac{2 q}{\sqrt{1-a^4r_0^4q^2}}\left.\left[ 
\int_{y_c}^{\Lambda} \frac{1 + a^4 r_0^4 y^4}{\sqrt{(y^4-1)(y^4-y_c^4)}} dy
- a^4 r_0^4 \Lambda\right]\right|_{\Lambda \to \infty}
\ee 
The above equation therefore gives us the quark antiquark separation
$L(q) = \ell(q)/(\pi T)$ of the dipole as a function of the constant 
of motion $q$.
It is not possible to perform the integration in \eqref{regellq} and give an
analytic expression of $\ell(q)$. So, we will perform the integration
numerically for different fixed values of the rapidity  $\eta$ and
the noncommutativity parameter $ar_0 \sim \hat \lambda^{1/4} T \sqrt{\theta}$
and use this to evaluate the quark-antiquark potential. However, we can give
the analytic expression of $\ell(q)$ and the screening length only for some
special values of the rapidity and the noncommutativity parameter which will be
discussed later.

Now substituting $y'$ from \eqref{yprime} into the action \eqref{ngtwo} we get,
\be\label{ngthree}
S(\ell) = \frac{{\cal T} T \sqrt{\hat \lambda}}{\sqrt{1-a^4r_0^4q^2}}
\int_{y_c}^{\Lambda} \frac{y^4 - \cosh^2\eta}{\sqrt{(y^4-1)(y^4-y_c^4)}} dy
\ee 
However, as it is known \cite{Liu:2006he} from the commutative case that 
this action is
divergent as we take $\Lambda \to \infty$. The reason for this divergence for
the commutative case is that this action contains the quark-antiquark
self-energy $S_0$ and once this is subtracted from $S(\ell)$, the action
becomes finite and from there one calculates the potential. In the
noncommutative case also, we need to subtract the quark-antiquark self energy
$S_0$. This is calculated by evaluating the Nambu-Goto action for the single
string stretched between the horizon $r_0$ and the boundary at $r_0\Lambda$
and then multiply by 2 (for the two strings associated with quark and 
antiquark). The Nambu-Goto action in this case is evaluated by considering a
quark moving in the $x^1$-direction (which is commutative) and so, the gauge 
condition one uses is
$\tau = t$, $\sigma = r$, $x^1 = x^1(\sigma)$, $x^2(\sigma) = x^3(\sigma) =$
constant\footnote{By constant here we mean that $x^2$ and $x^3$ are not
functions of $r=\sigma$. However since these are the noncommutative directions
they will have the expected fuzziness in the $x^2$-$x^3$ plane.}. 
However, since $x^2$ and $x^3$ are the noncommutative directions the
computation of $S_0$ remains unaffected by the noncommutativity. Therefore,
$S_0$ takes exactly the same form as evaluated in the commutative case given as
\cite{Liu:2006he}
\be\label{szero}
S_0 = {\cal T} T \sqrt{\hat \lambda}\int_1^\Lambda dy
\ee   
So, subtracting $S_0$ from $S(\ell)$ we get,
\bea\label{sminusszero}
S(\ell) - S_0 &=& \frac{{\cal T}T \sqrt{\hat \lambda}}{\sqrt{1-a^4r_0^4q^2}}
\left[\int_{y_c}^{\Lambda} dy \left\{\frac{y^4 -
      \cosh^2\eta}{\sqrt{(y^4-1)(y^4-y_c^4)}} - \sqrt{1 - a^4 r_0^4
      q^2}\right\} \right.\nn
& & \qquad\qquad\qquad\qquad\qquad\qquad\qquad\left. - \sqrt{1-a^4 r_0^4 q^2}
(y_c-1)\right]
\eea        
It is clear from \eqref{sminusszero} that in the commutative case when
$a^2 r_0^2 \sim \theta$ is put to zero, $S(\ell) -S_0$ is indeed finite 
when we take $\Lambda \to \infty$. However, this is not the case when
noncommutativity is present. The reason is, as we mentioned before, the
NCYM theory does not live at $\Lambda = \infty$, but at a finite $\Lambda$
\cite{Maldacena:1999mh,Dhar:2000nj}. (Another way of understanding this 
divergence has been mentioned earlier from the arguments given in 
\cite{Alishahiha:1999ci,Bigatti:1999iz,Haque:2009hz}.) 
So, we will subtract the divergent term from \eqref{sminusszero} and then take
$\Lambda \to \infty$. Doing that we find the finite quark-antiquark potential
as,
\bea\label{qqbar}
E(\ell) &=& \frac{S-S_0-S_{\rm div}}{\cal T}\nn
&=& \frac{T \sqrt{\hat \lambda}}{\sqrt{1-a^4r_0^4q^2}}
\left[\int_{y_c}^{\Lambda} dy \left\{\frac{y^4 -
      \cosh^2\eta}{\sqrt{(y^4-1)(y^4-y_c^4)}} - \sqrt{1 - a^4 r_0^4
      q^2}\right\} \right.\nn
& & \qquad\qquad\qquad \left.\left. - \sqrt{1-a^4r_0^4q^2}(y_c-1)
- \left(1 - \sqrt{1-a^4r_0^4q^2}\right)\Lambda\right]\right|_{\Lambda \to
\infty}
\eea  
where in the above $S_{\rm div}$ is $(1-\sqrt{1-a^4r_0^4q^2})\Lambda$ with 
$\Lambda \to \infty$.   
Here also it is not possible to perform the integration
in \eqref{qqbar} in a closed form. So, we will obtain the quark-antiquark
potential numerically. We first plot $\ell(q)$ vs. $q$ using \eqref{regellq}
and use it to plot $E(\ell)$ vs. $\ell$ from \eqref{qqbar} at different fixed
values of the rapidity $\eta$ and the noncommutativity parameter
$ar_0$. 

\subsection{Plots and discussion of the results}

In this subsection we give and discuss the various plots of quark-antiquark 
separation $\ell(q)$
as a function of constant of motion $q$ and the velocity dependent 
quark-antiquark
potential $E(\ell)$ as a function of the quark-antiquark separation length 
$\ell$ for various values of the rapidity ($\eta$) as well as the 
noncommutativity parameter ($ar_0 \sim \sqrt{\theta}$). 

\begin{figure} [ht]
\begin{center}
\subfigure[]{
\includegraphics[scale=0.3, angle=-90]{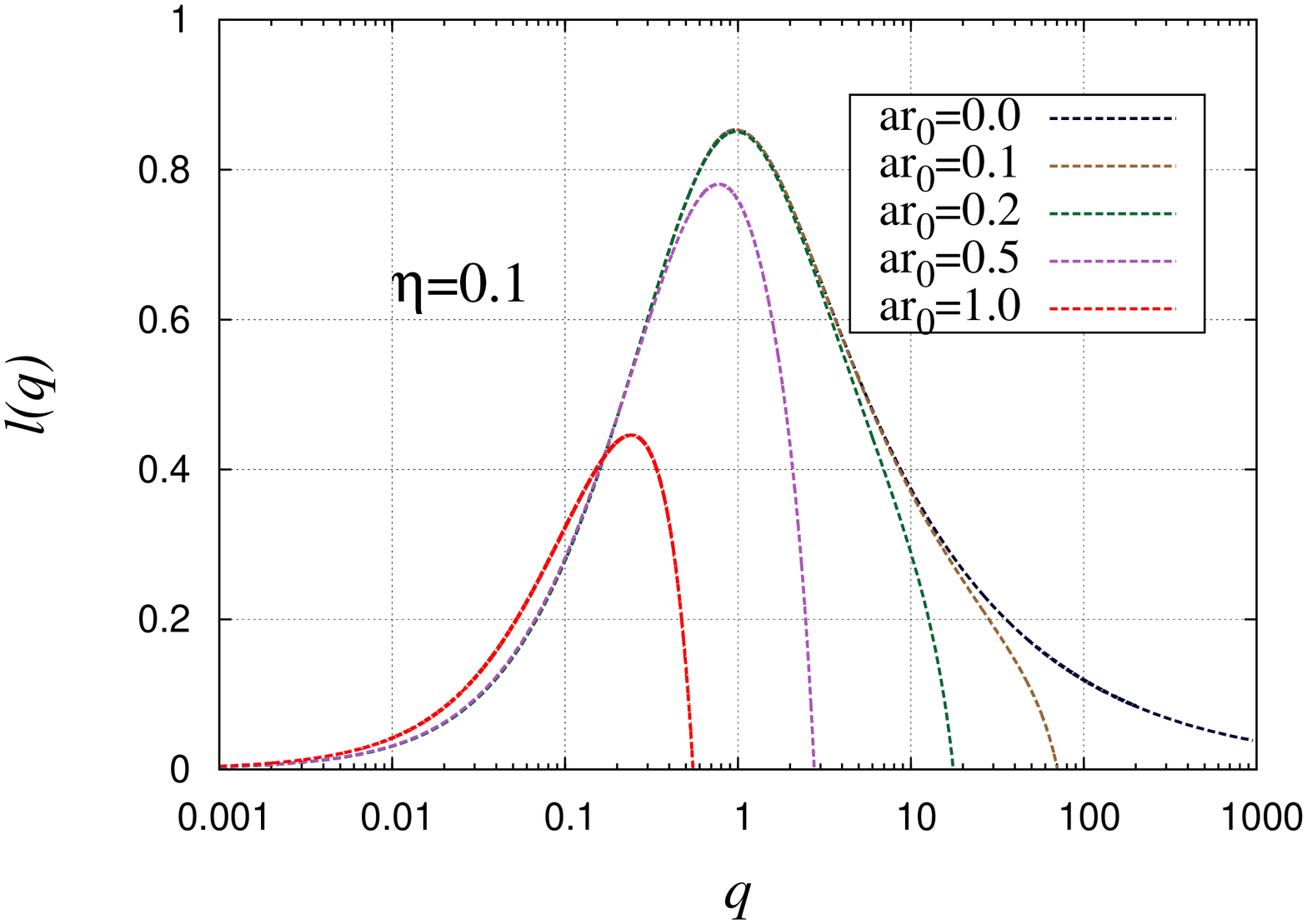}}
\subfigure[]{
\includegraphics[scale=0.3, angle=-90]{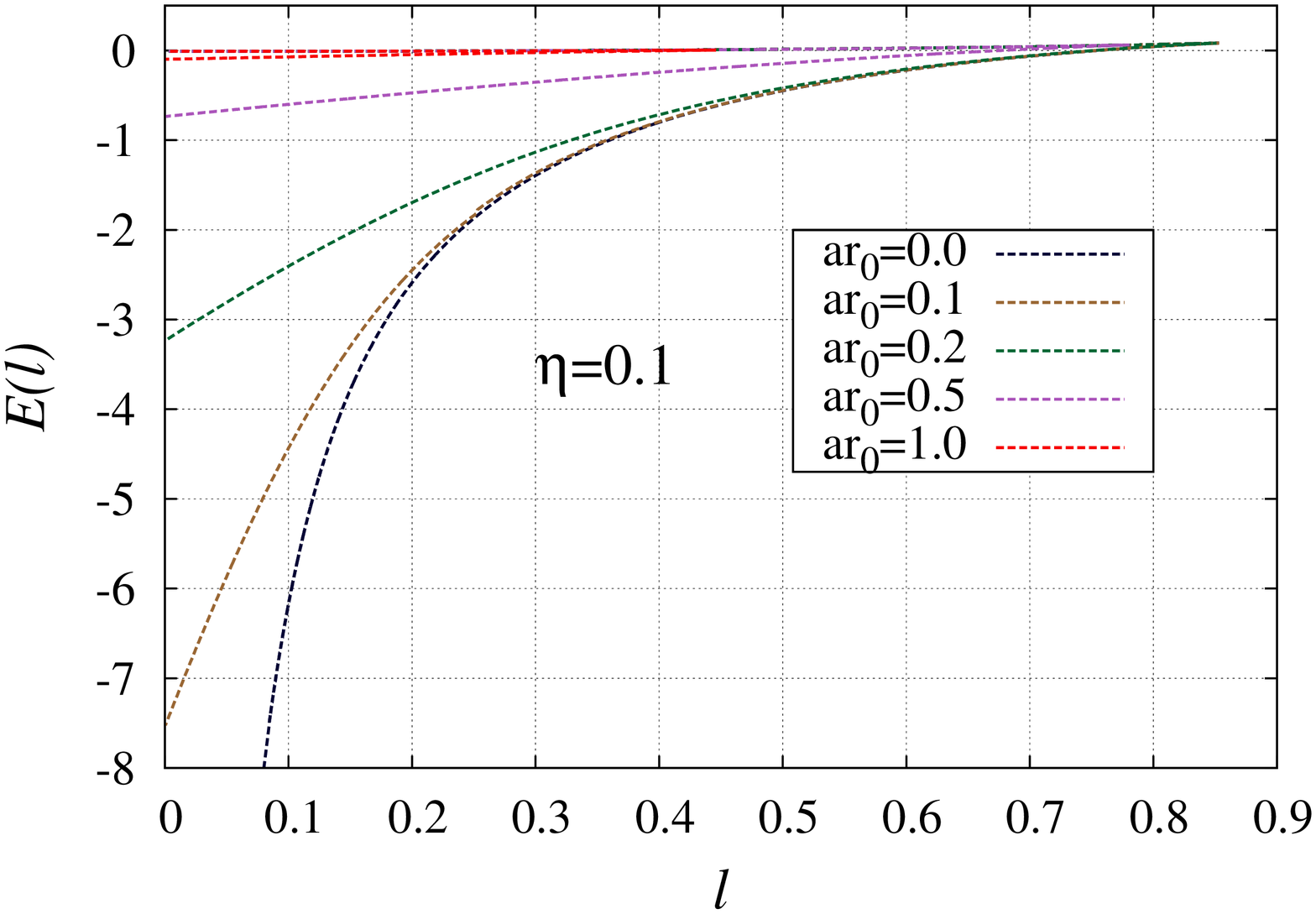}}
\caption{\label{fig1} \small{(a) shows the plot of the
quark-antiquark
separation $\ell(q)$ as a function of the const. of motion $q$
for different values of the noncommutativity parameter $ar_0$, when the rapidity
is kept fixed at $\eta = 0.1$. (b) shows the plot
of the properly normalized quark-antiquark potential $E(\ell)$ as a function 
of $\ell$ for the same set of values of the noncommutativity parameter
with the same value of $\eta=0.1$.}}
\end{center}
\end{figure}

\begin{figure} [ht]
\begin{center}
\subfigure[]{
\includegraphics[scale=0.3, angle=-90]{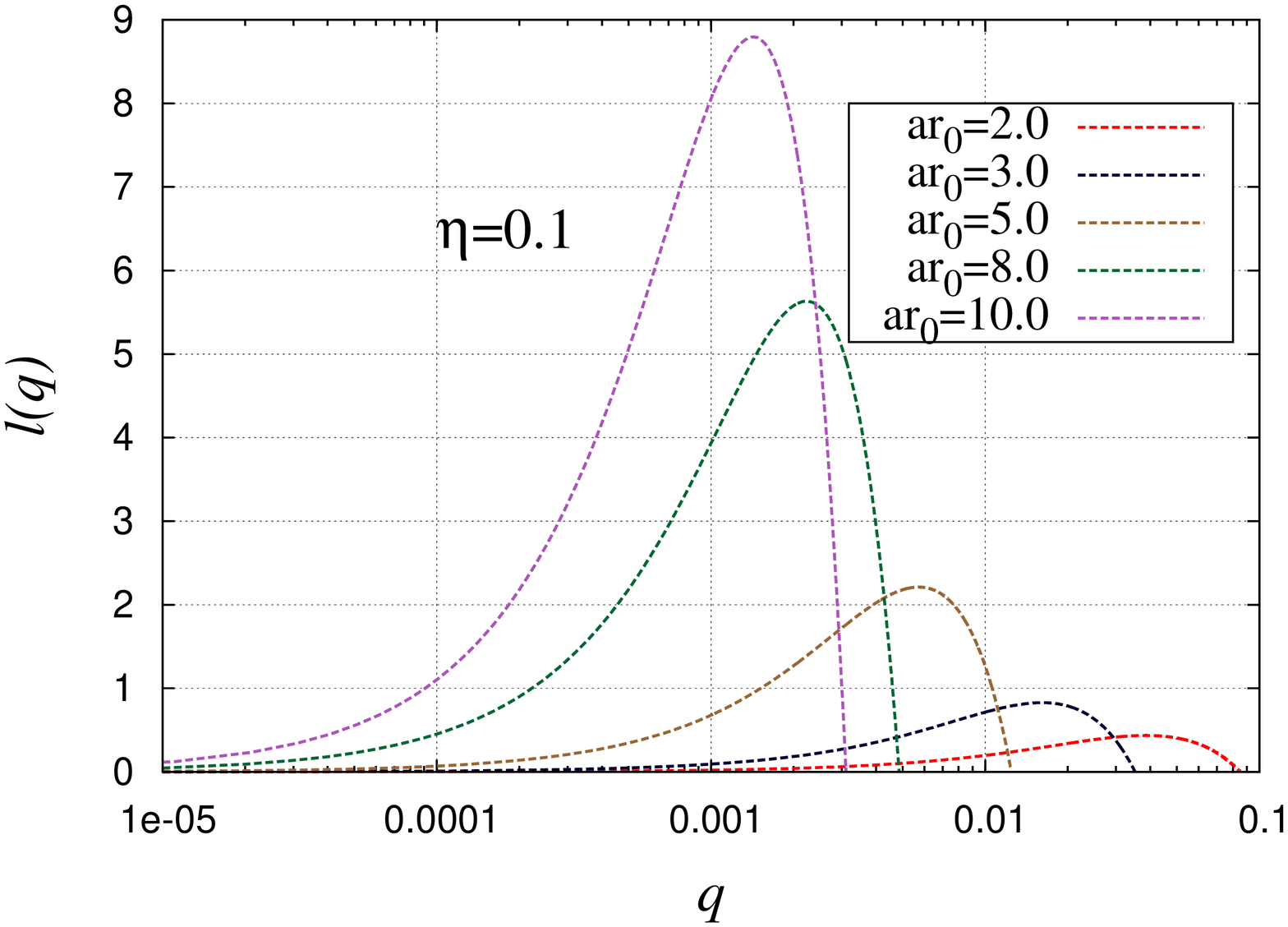}}
\subfigure[]{
\includegraphics[scale=0.3, angle=-90]{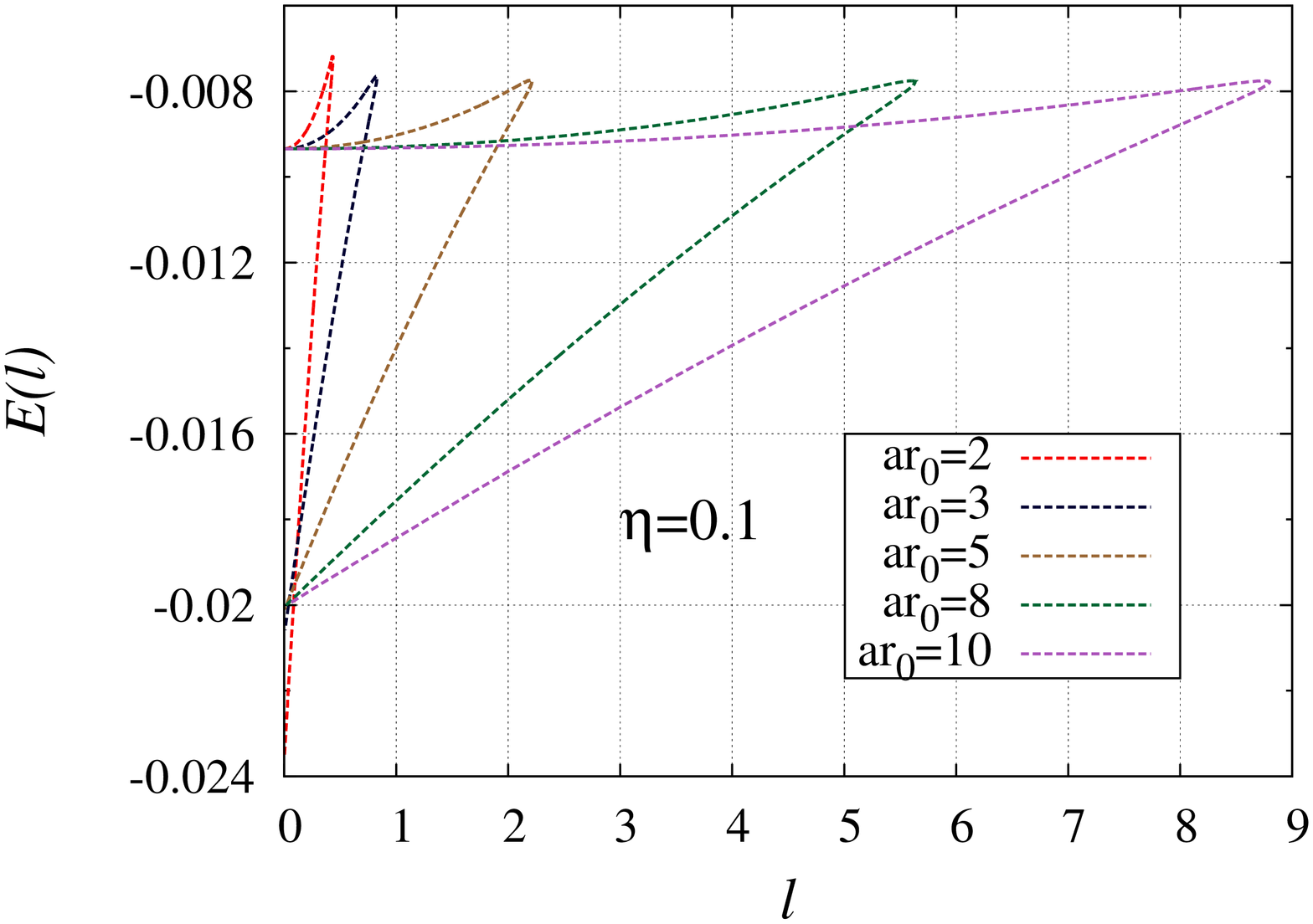}}
\caption{\label{fig2} \small{(a) shows the plot of the
quark-antiquark
separation $\ell(q)$ as a function of the const. of motion $q$ for
different values (but larger than those given in Figure 1) of the 
noncommutativity 
parameter $ar_0$ with the same $\eta = 0.1$ as in Figure 1. (b) shows the plot
of the properly normalized quark-antiquark potential $E(\ell)$ as a 
function of $\ell$ for
the same set of values of the noncommutativity parameter with the 
same $\eta=0.1$.}}
\end{center}
\end{figure}

Note that each figure contains two parts (a) and
(b). While in (a) we plot $\ell(q)$ vs $q$, in (b) we plot the quark-antiquark 
potential $E(\ell)$ vs $\ell$. In plot (b) we make use of plot (a) to obtain 
$q$ for each value of $\ell(q)$. Generically in all these figures the
$\ell(q)$ in part (a) starts at zero for $q=0$ (in fact for small $q$, 
$\ell(q) \to 0$ as $q$ which can be seen from \eqref{regellq}) and as 
$q$ increases it ends at zero either at a finite value of $q$, denoted as
$q_{\rm max}$,  
(when noncommutativity is present or $ar_0 \neq 0$) or at $q \to \infty$ (when
there is no noncommutativity or $ar_0 = 0$) 
(for large $q$, $\ell(q) \to 0$ as $q^{-1/2}$ which can be seen again from 
\eqref{regellq}). In between, $\ell(q)$
has a single maximum at some finite $q<q_{\rm max}$ indicating that there 
exists a
screening length (proportional to $\ell_{\rm max}$) beyond which there is 
no solution to   
\eqref{regellq}. So, given a curve, there is an $\ell_{\rm max}$ beyond which
there is no dipole solution, that is, the dipole dissociates and below 
$\ell_{\rm
  max}$ there are two dipoles at a fixed $\ell$ for two different values of 
$q$.  Accordingly, $E(\ell)$ or $Q$-${\bar Q}$ potential in part (b) 
generically 
in all the figures have two branches corresponding to the dipoles with 
two different values of $q$. The smaller value of $q$ corresponds to the upper
branch and the dipole has higher energy, whereas, the larger value of $q$ 
corresponds to the lower branch and the dipole has lower energy. So, the
dipole with lower $q$ will be metastable and will go to the state with
higher $q$ as it is energetically more favorable. 

 \begin{figure} [ht]
\begin{center}
\subfigure[]{
\includegraphics[scale=0.3, angle=-90]{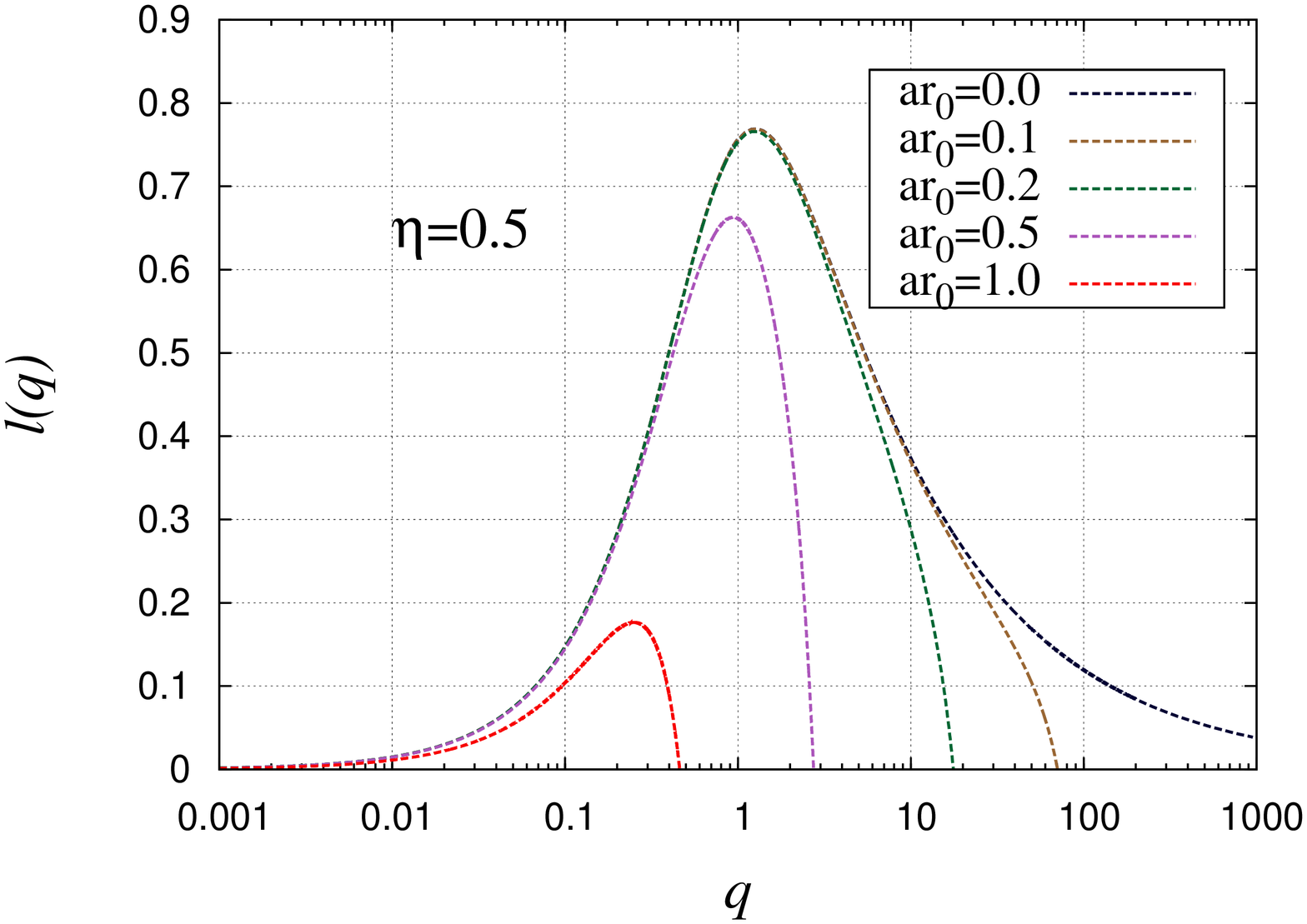}}
\subfigure[]{
\includegraphics[scale=0.3, angle=-90]{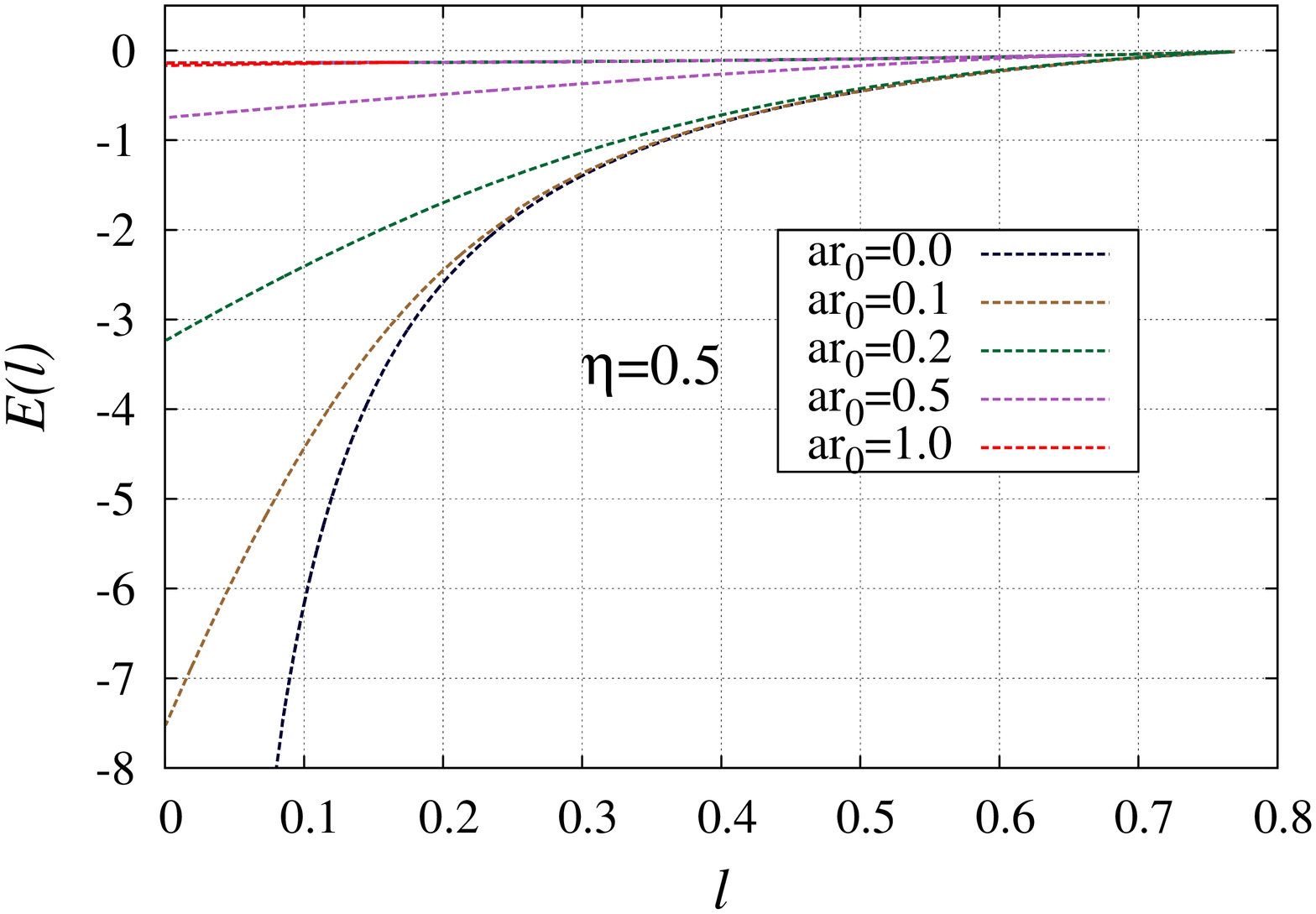}}
\caption{\label{fig3} \small{(a) shows the plot of the
quark-antiquark
separation $\ell(q)$ as a function of the const. of motion $q$ for
different values of the noncommutativity parameter $ar_0$, but now the
rapidity is kept fixed at $\eta = 0.5$. (b) shows the plot
of the properly normalized quark-antiquark potential $E(\ell)$ 
as a function of $\ell$ for the same set of values of the  noncommutativity 
parameter with the same value of $\eta = 0.5$}}
\end{center}
\end{figure}

\begin{figure} [ht]
\begin{center}
\subfigure[]{
\includegraphics[scale=0.3, angle=-90]{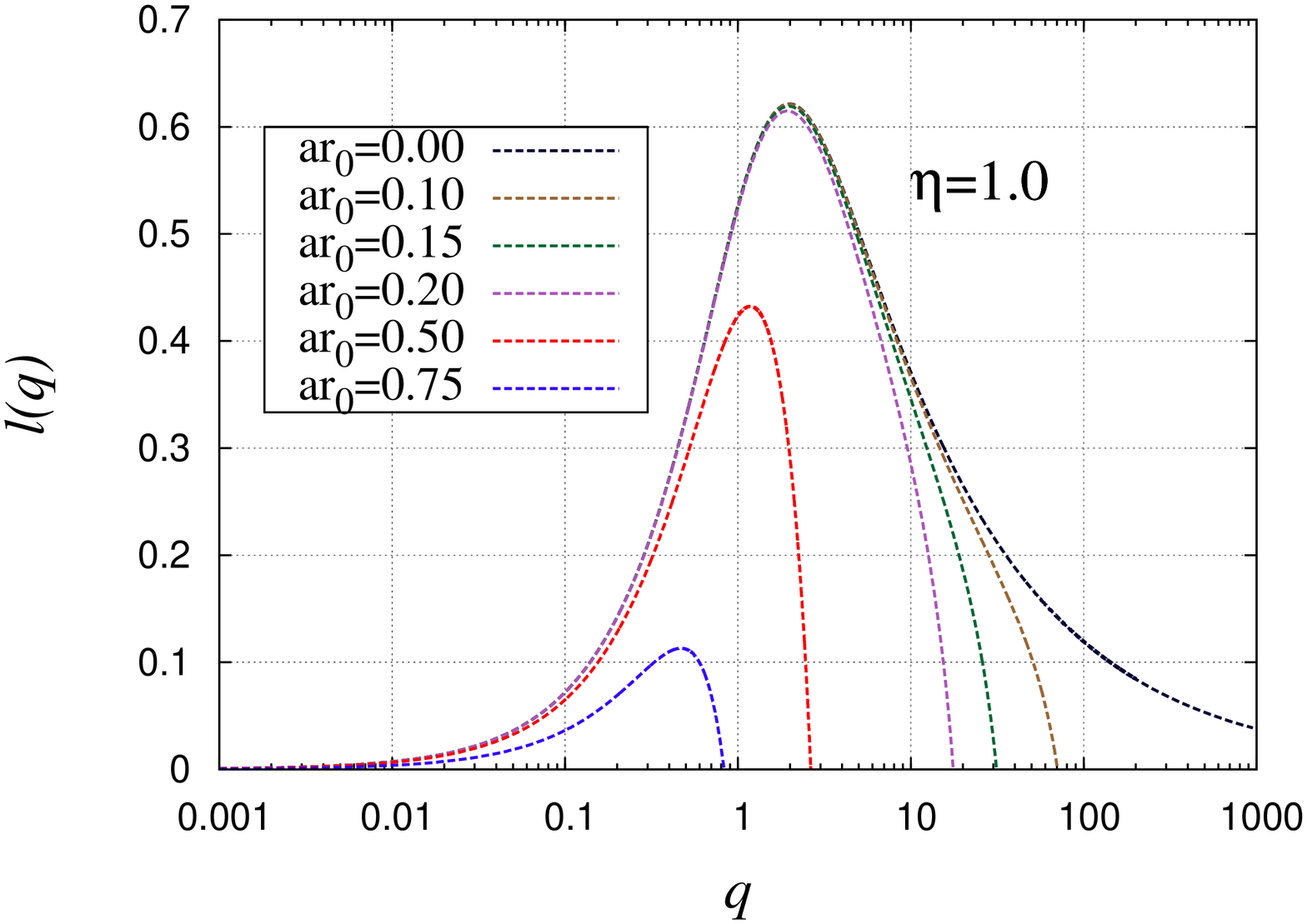}}
\subfigure[]{
\includegraphics[scale=0.3, angle=-90]{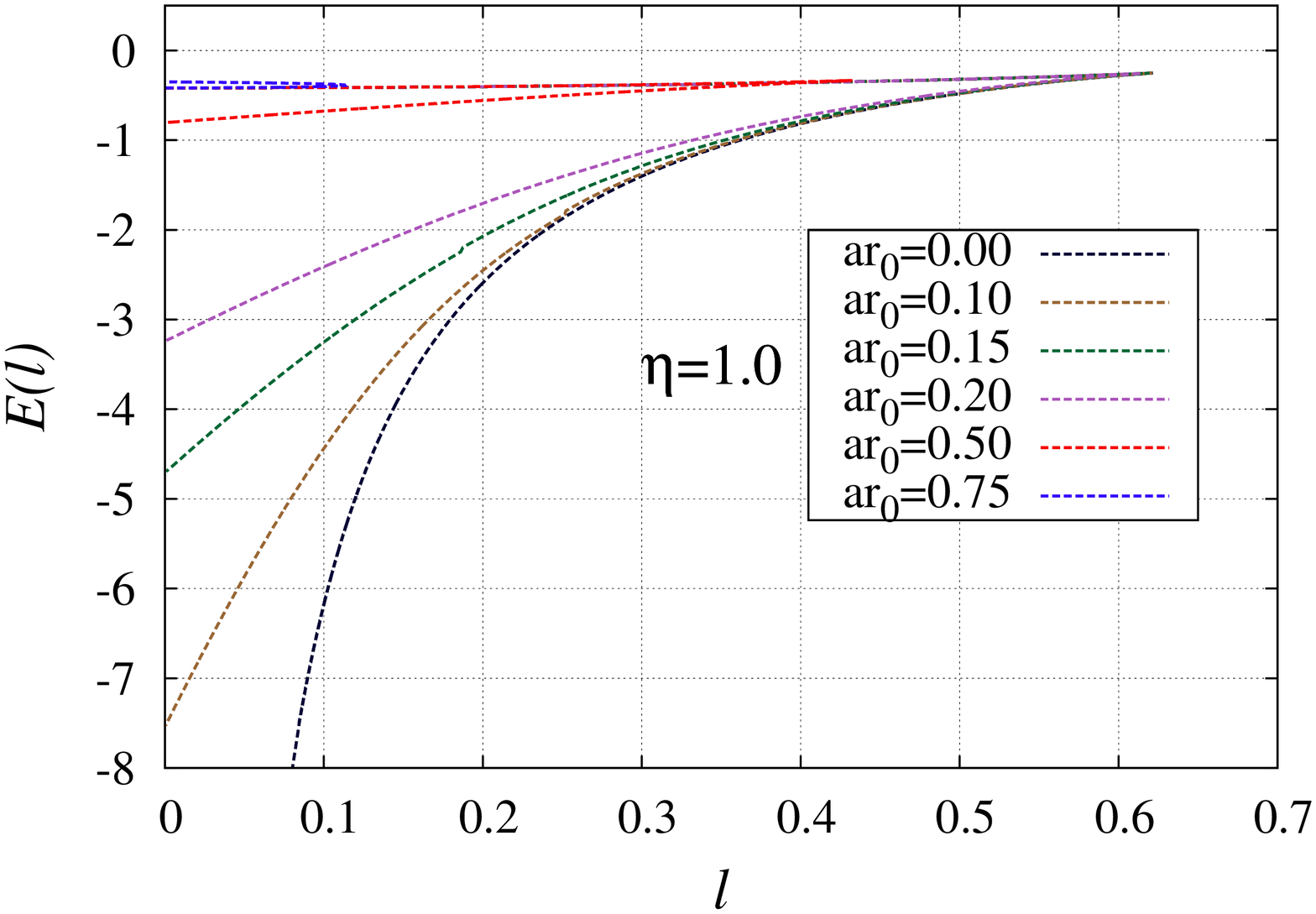}}
\caption{\label{fig4} \small{(a) shows the plot of the
quark-antiquark
separation $\ell(q)$ as a function of the const. of motion $q$ for
different values of the noncommutativity parameter $ar_0$, but now the
rapidity is fixed at $\eta=1.0$. (b) shows the plot
of the properly normalized quark-antiquark potential $E(\ell)$ as a 
function of $\ell$ for the same set of values of the noncommutative 
parameter with the same $\eta = 1.0$.}}
\end{center}
\end{figure} 

Also from part (b) we note that there exists a critical value of the rapidity
$\eta_c$ (whose value changes with the value of the
noncommutativity parameter), beyond which the entire upper branch of the
$E(\ell)$ curve is negative. However, below this value, i.e., $\eta < \eta_c$,
the $E(\ell)$ curve crosses zero at $\ell = \ell_c$, continues to rise till
$\ell = \ell_{\rm max}$ and then turns back crossing zero again at $\ell =
\ell_c' > \ell_c$. There are various portions of the $E(\ell)$ vs $\ell$ curve,
namely, $\ell < \ell_c$, $\ell=\ell_c$, $\ell_c < \ell < \ell_c'$, $\ell =
\ell_c'$, $\ell_c' < \ell < \ell_{\rm max}$ and $\ell>\ell_{\rm max}$ which
need separate
discussions as far as the quark-antiquark potential is concerned. However, the
discussion is pretty much the same as in the commutative case given in 
\cite{Liu:2006he, Chakraborty:2011ah} and will not be repeated here. We 
will point out the differences due to noncommutativity as we go along.   

\begin{figure} [ht]
\begin{center}
\subfigure[]{
\includegraphics[scale=0.3, angle=-90]{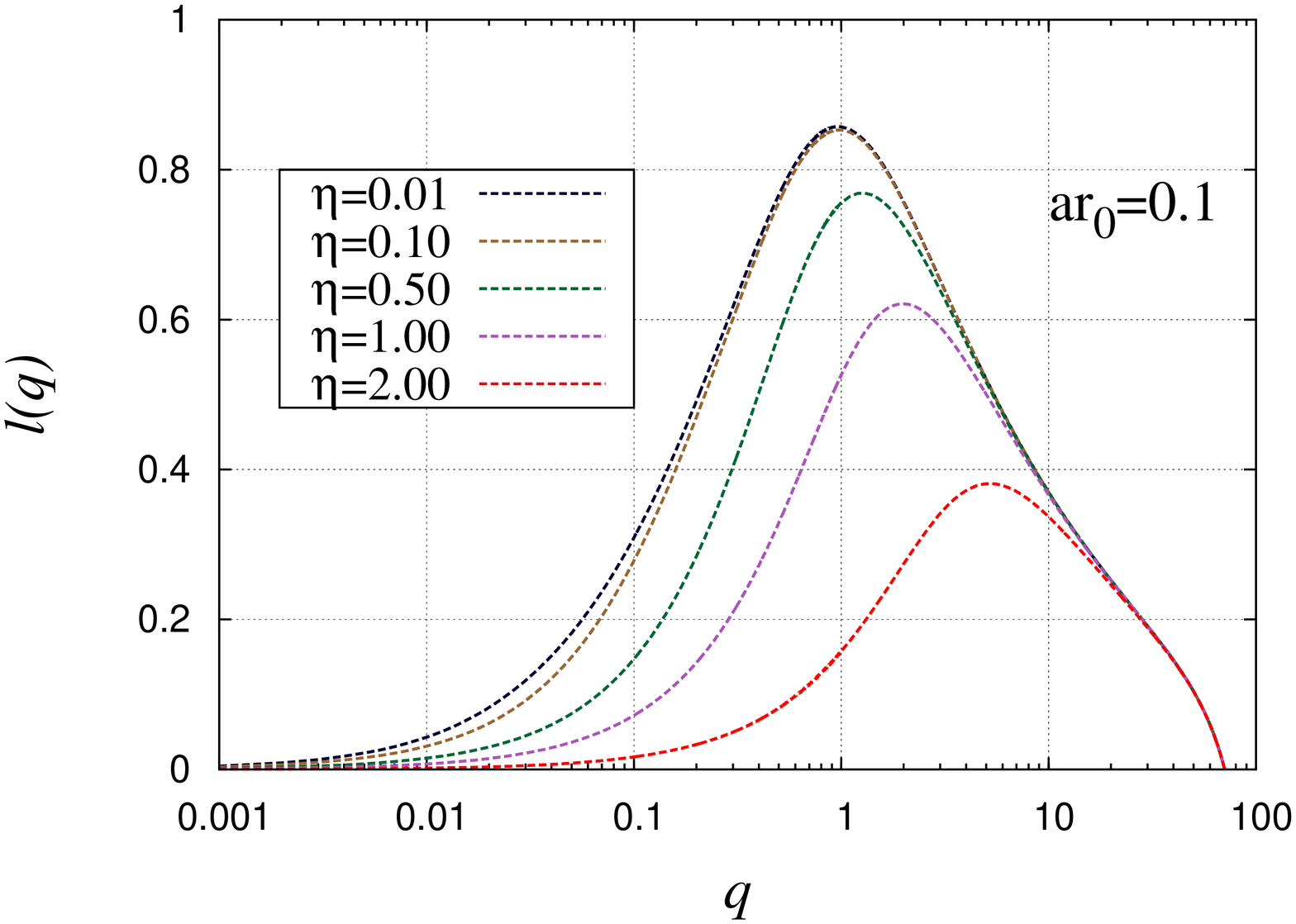}}
\subfigure[]{
\includegraphics[scale=0.3, angle=-90]{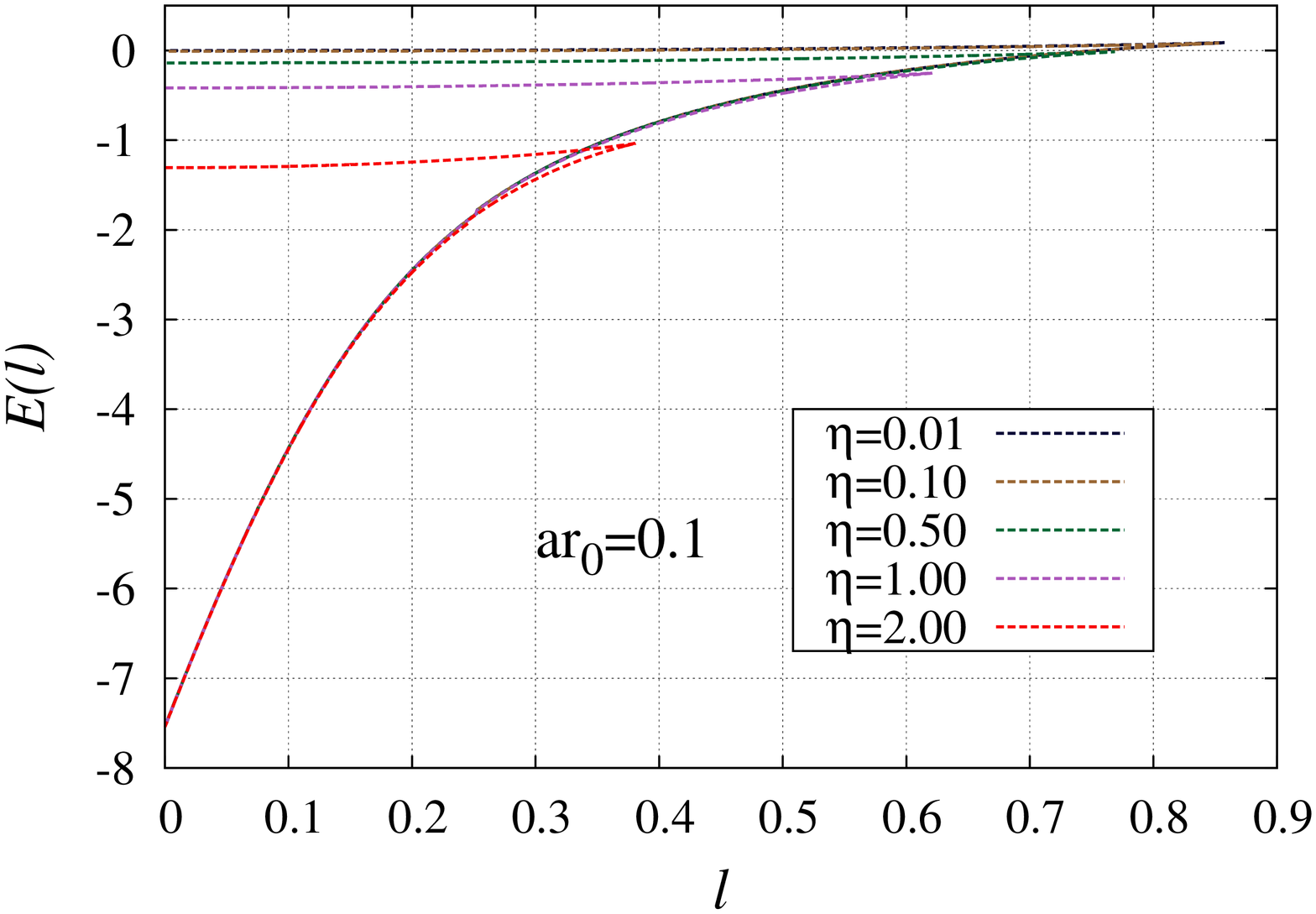}}
\caption{\label{fig5} \small{(a) shows the plot of 
quark-antiquark
separation $\ell(q)$ as a function of the const. of motion $q$ for
different values of the rapidity $\eta$ with the noncommutativity
parameter kept fixed at $ar_0=0.1$. (b) shows the plot
of the properly normalized quark-antiquark potential $E(\ell)$ as a 
function of $\ell$ for the same set of values of the rapidity 
with the same $ar_0=0.1$.}}
\end{center}
\end{figure}

\begin{figure} [ht]
\begin{center}
\subfigure[]{
\includegraphics[scale=0.3, angle=-90]{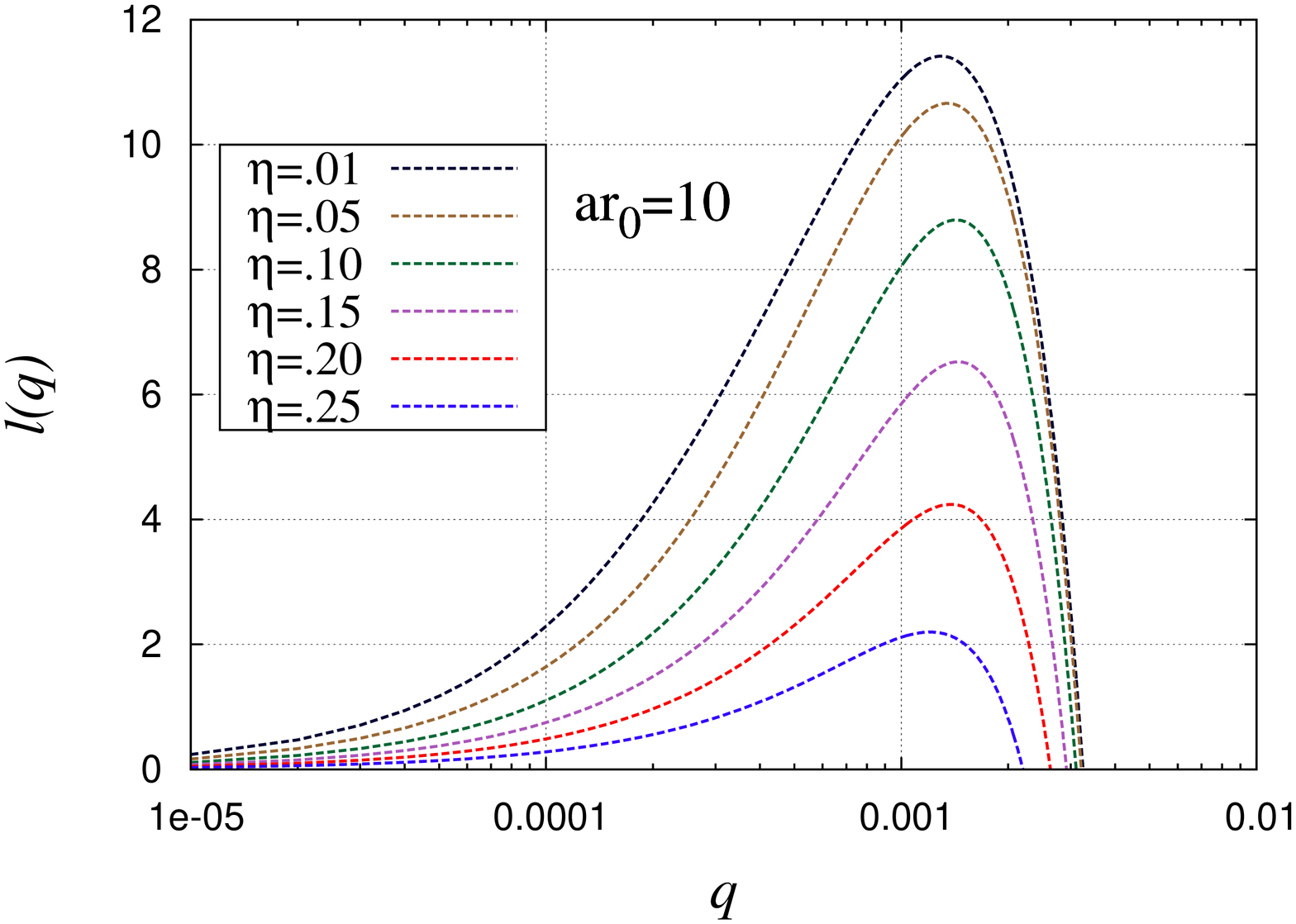}}
\subfigure[]{
\includegraphics[scale=0.3, angle=-90]{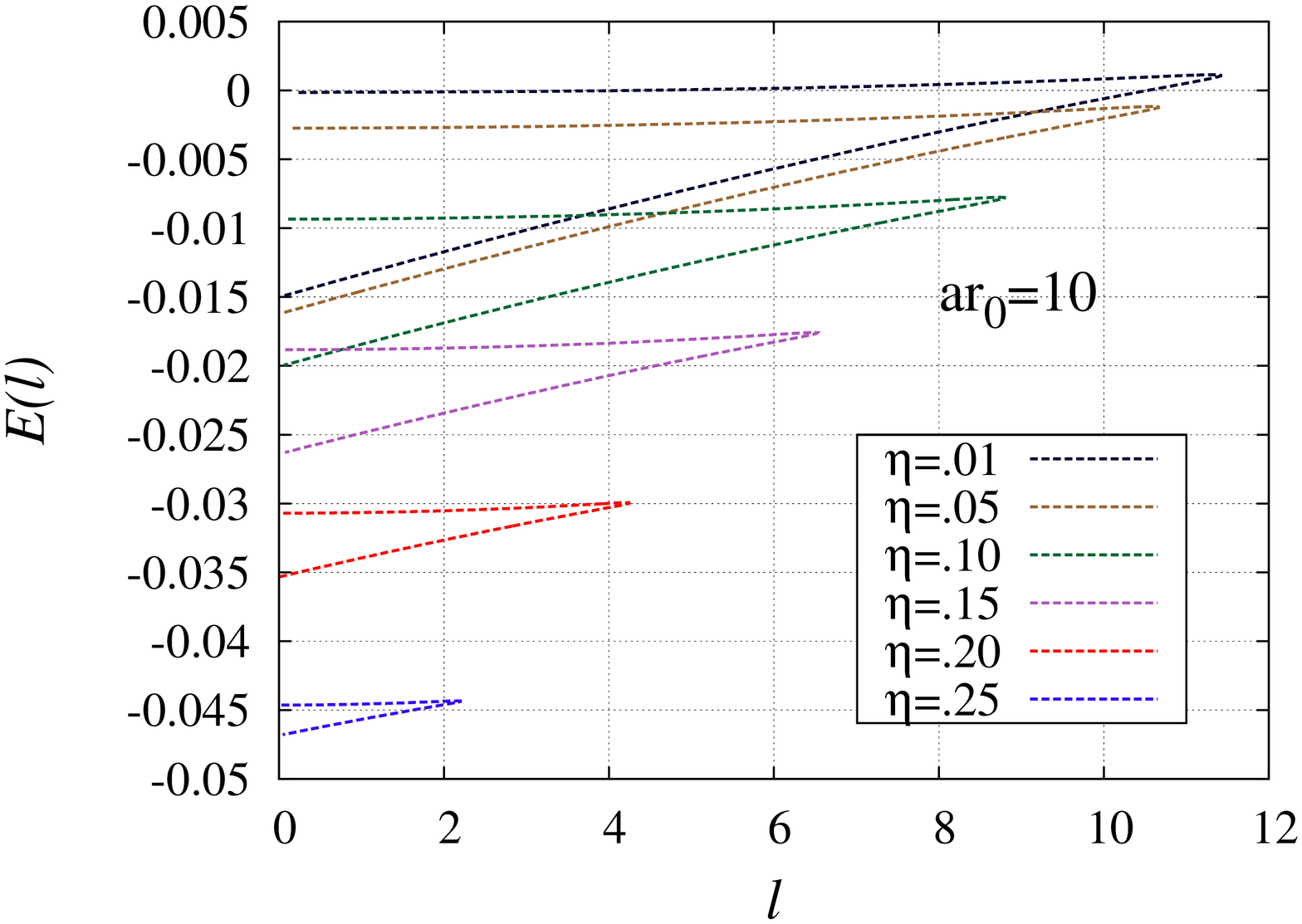}}
\caption{\label{fig6} \small{(a) shows the plot of the
quark-antiquark
separation $\ell(q)$ as a function of the const. of motion $q$ for
different values of the rapidity  $\eta$ but now
the noncommutativity parameter is kept fixed at a much larger value
$ar_0=10.0$. (b) shows the plot
of the properly normalized quark-antiquark potential $E(\ell)$ as a 
function of $\ell$ for the same set of values of the rapidity
with the same $ar_0=10.0$.}}
\end{center}
\end{figure}

In Figures 1 and 2, the rapidity is fixed at $\eta = 0.1$. However,
in Figure 1 the noncommutativity parameter $ar_0$ takes small values starting 
from $0.0$ (where there is no noncommutativity) to $1.0$, whereas, in Figure
2 the noncommutativity parameter takes fairly large values starting from
$2.0$ to $10.0$. The main difference between the commutative results and the
noncommutative results is that in the former case the constant of motion $q$
can take arbitrarily large values, but in the latter case $q$ can not exceed
certain finite value ($q_{\rm max}$) because beyond this value the 
quark-antiquark separation $\ell(q)$ becomes negative which is unphysical. The
reason behind this cut-off is the regularization of the integral made in
\eqref{regellq}. Note that the last term in \eqref{regellq} is subtracted in
order to make $\ell(q)$ finite as $\Lambda \to \infty$. However, as $q$
increases, $y_c$ increases which makes the last term dominate over the integral
and therefore $\ell(q)$ becomes negative. Thus this effect is due to the
noncommutativity of the underlying boundary theory. We see from Figure 1(a)
that as the noncommutativity
parameter $ar_0$ increases, $\ell(q)$ curve deviates more and more from the 
commutative curve, the $\ell_{\rm max}$ falls and the peak shifts towards the 
left (i.e., the maximum occurs at a smaller value of $q$). In particular,
the deviation from the commutative case becomes more pronounced after 
$\ell_{\rm max}$ is reached. However this feature
continues upto certain value of $ar_0 \sim 2.0$ and as it is increased 
further (see Figure 2(a)) the
$\ell(q)$ curve now deviates more from the commutative case throughout
the allowed range of $q$, but the maximum
value, $\ell_{\rm max}$, again starts rising and the peak as before 
shifts further towards left,
i.e., towards smaller values of $q$. Note that the scales in Figures 1 
and 2 are quite
different and are chosen so such that we can see the plots for various higher 
values of $ar_0$. Figure 1(b) and 2(b) shows the plot of the velocity
dependent $Q$-$\bar Q$ potential $E(\ell)$ vs the quark-antiquark separation 
length $\ell$ for $\eta=0.1$ with various values of the noncommutativity
parameter $ar_0$. As we mentioned before each curve in this case has two
branches corresponding to the two dipole solutions obtained in Figures (a).
The slight deviation of $\ell(q)$ from the commutative case for small values of
$q$, i.e., below the value of $q$ corresponding to $\ell(q) = \ell_{\rm max}$,
(see Figure 1(a)) is reflected in that fact that in Figure 1(b) the upper 
branches almost merge with the commutative counterpart whereas the greater
deviation in $\ell(q)$ after $\ell_{\rm max}$ is reached leads to a rise
in the lower branch of the $E(\ell)$ curve from the commutative case in
Figure 1(b). However, as the noncommutativity parameter is increased 
the overall deviation (particularly in the lower branch) of the $E(\ell)$ 
curve is more pronounced from its commutative value.   
In contrast, in Figure 2(b) as the noncommutativity parameter 
is further increased, $E(\ell)$, in general, dips slightly for both the 
branches.

The feature that the screening length ($\sim \ell_{\rm max}$) initially drops and
then rises and correspondingly the lower branch of the potential $E(\ell)$
rises and then drops as we go on increasing the noncommutativity parameter
$ar_0$ (with the transition occurring at around $ar_0=2.0$), occurs only for
the smaller value of the rapidity, $\eta=0.1$. As the rapidity
becomes higher its effect starts to dominate and the transition (from falling
$\ell_{\rm max}$ to rising $\ell_{\rm max}$ as the noncommutativity
parameter is increased) is suppressed so that now 
the screening length continuously drops and the
lower branch of the $Q$-$\bar Q$ potential continuously rises. We have seen
this to happen for $\eta = 0.5$ and $\eta=1.0$. That is why we have given
those plots only for the smaller values of $ar_0$ in Figures 3 and 4. Although
the details of these plots are different, the general features remain very 
similar to those of $\eta = 0.1$ as we have discussed and so, we do not 
elaborate their discussion further to avoid repetitions.   

We have pointed out that with different rapidities the general features of 
the plots 1, 3 and 4 (i.e., for small noncommutativity) are quite similar 
although the details are different. We see from plots (a) that in all three 
cases, the
maximum of the $\ell (q)$ curve ($\ell_{\rm max}$) or the screening length 
drops as 
the noncommutativity is increased. This implies that with the increase of
noncommutativity, less and less number of dipoles will be formed in the QGP
and there 
will be more $J/\Psi$ suppression \cite{Digal:2001iu}. On the other hand, 
from plots (b) we see that with the
increase of noncommutativity the $Q$-$\bar Q$ potential rises (see 
the lower curves which correspond to the stable states) in value which means 
that the quark and antiquark will be 
more and more loosely bound and eventually there will be no bound state
formation. This may be expected since there is a fuzziness in the direction 
of the dipole due to noncommutativity. However, in plot 2(a), 
when the noncommutativity is large (and the rapidity remains small) we see 
that around $ar_0=2.0$, the maximum
of the $\ell(q)$ curve ($\ell_{\rm max}$) starts rising again and so more dipoles
can form, but from plot (b) we see (from the lower curve) that in this case the
quarks and antiquarks will be very very loosely bound. This does not happen 
when
the rapidity is large (we have not shown the plots for this case with large
noncommutativity). Here the screening length ($\sim \ell_{\rm max}$) 
continually drops as we have checked. This seems to indicate that the 
effect of noncommutativity (for large noncommutativity) gets suppressed 
when the rapidity or the velocity of the dipole is high. But for small 
noncommutativity the 
effect is quite prominent even when the velocity of the dipole is high 
(as shown in plots 3, 4).    
   
In contrast to Figures 1 -- 4, where we plot $\ell(q)$
vs $q$ and $E(\ell)$ vs $\ell$ for fixed value of the rapidity
$\eta$ but with varying values of noncommutativity parameter $ar_0$, in
Figures 5, 6, we plot the same functions for fixed value of noncommutativity 
parameter $ar_0$, but varying values of rapidity  $\eta$. In Figure 
5, $ar_0$ is fixed to a small value $0.1$ whereas in Figure 6, it is fixed 
to a large value $10.0$. In Figure 5(a) we find that as the rapidity 
increases the screening length decreases (which means there will be less
dipole formation i.e., more $J/\Psi$ suppression) and the peaks shift 
towards right,
i.e., to the larger value of $q$. This is expected as in the commutative case
also there is a decrease in screening length with the increase in rapidity. 
Further note that for larger value of $q$, $\ell(q)$ 
becomes independent of $q$. This is also shown in $E(\ell)$ vs $\ell$ plot in
Figure 5(b), i.e. the lower branches of the curves merge. Contrast this
to the case when $ar_0$ is changed (but still kept small) keeping $\eta$
fixed, when the lower part of the $\ell(q)$ curve (i.e., small $q$) does not
exhibit significant deviation, the peak shifts towards left and the
upper branch of the $E(\ell)$ curves merge. So we can think of $\eta$ and 
$ar_0$ as sort of having opposite effects. 
These features are
also evident for large values of $ar_0$ given in Figure 6. In figure 6(a) as
the rapidity $\eta$ increases the screening length decreases and the
peaks shift towards right, but since now the scale of $q$-axis is very much
enlarged this is not much visible (this is also due to the fact that $\eta$
now changes by a very small amount). The independence of $\ell(q)$ with $q$
for larger values of $q$ is not evident in this case due to the  
differences in scale in $q$-axis for Figure 5(a) and Figure 6(a), however, it
is clear that it has this tendency. Unlike in Figure 5(b) the lower branches of
the $E(\ell)$ curves do not merge for different values of $\eta$ as is
evident from Figure 6(b). However, the spread is again due to the enlarged
(compared to Figure 5(b)) scale of the $E(\ell)$ axis which is chosen to
show the two branches of the $E(\ell)$ curve distinctly.    

\subsection{Screening length in a special case}

The expression for the regularized quark-antiquark separation length $\ell(q)$
as a function of the constant of motion $q$ is given
in \eqref{regellq}. However, as we have mentioned, it is not possible to
perform the integration occurring there in a closed form in general and give
an exact analytic expression for $\ell(q)$. That is why we have numerically
solved \eqref{regellq} and plotted $\ell(q)$ in the previous subsection. This
has nothing to do with the noncommutativity of the underlying gauge theory and
this happens also for the case of commutative theory. For the case of
commutative theory it is possible to give an exact analytic expression of
$\ell(q)$ only in the large velocity or large rapidity limit. Noncommutativity,
on the other hand, makes the analysis a little bit more involved and in 
this case
it is possible to obtain the analytic expression only when the rapidity is
large and the noncommutativity is small with the product remaining small. For
large $\eta$ or large $y_c$, the expression for $\ell(q)$ in \eqref{regellq}    
can be expanded as follows,
\bea\label{expandell}
\ell(q) &=& \left[\frac{2q}{\sqrt{1-a^4r_0^4q^2}}\int_{y_c}^\Lambda \frac{
1+a^4r_0^4y^4}{y^2\sqrt{y^4-y_c^4}} dy + \frac{q}{\sqrt{1-a^4r_0^4q^2}}
\int_{y_c}^\Lambda \frac{1+a^4r_0^4y^4}{y^6\sqrt{y^4-y_c^4}} dy\right.\nn
& & \qquad\qquad \left.\left.+ \frac{3q}{4\sqrt{1-a^4r_0^4q^2}}
\int_{y_c}^\Lambda \frac{
1+a^4r_0^4y^4}{y^{10}\sqrt{y^4-y_c^4}} dy + \cdots - \frac{2q
a^4r_0^4}{\sqrt{1-a^4r_0^4q^2}}\Lambda\right]\right|_{\Lambda \to \infty}
\eea
When $\Lambda \to \infty$, the above integrals can be evaluated and $\ell(q)$
can be written as a series expansion in inverse powers of $y_c$ as,
\be\label{ellqapprox}
\ell(q) =
\frac{q\sqrt{\pi}y_c}{\sqrt{1-a^4r_0^4q^2}}\left[-2a^4r_0^4
\frac{\Gamma\left(\frac{3}{4}\right)}{\Gamma\left(\frac{1}{4}\right)}   
+ (2+a^4r_0^4) \frac{\Gamma\left(\frac{3}{4}\right)}
{\Gamma\left(\frac{1}{4}\right)}\frac{1}{y_c^4} +
\left(1+\frac{3}{4}a^4r_0^4\right) \frac{\Gamma\left(\frac{7}{4}\right)}
{4\Gamma\left(\frac{9}{4}\right)}\frac{1}{y_c^8} + \cdots\right]
\ee
By construction the divergent last term in eq.\eqref{expandell} gets cancelled
with the divergent term in the first integral when $\Lambda \to \infty$. The
other integrals are convergent and makes the expression for $\ell(q)$ finite.
By taking the first three terms in the series we can obtain the value of $q$
and $y_c$ which maximizes $\ell(q)$ as,
\bea\label{qyc}
q^2 &=& 2\cosh^2\eta(1-15a^4r_0^4\cosh^2\eta)\nn
y_c^4 &=& \frac{\cosh^2\eta + q^2}{(1-a^4r_0^4q^2)}\,\,=\,\,
  3\cosh^2\eta(1-8a^4r_0^4 \cosh^2\eta)
\eea
In obtaining the above expressions we have assumed $a^4r_0^4 \ll 1$ and
$a^4r_0^4 \cosh^2\eta \ll 1$. Using \eqref{qyc} we obtain the maximum value of
$\ell$ upto next to leading order as,
\bea\label{lmax}
\ell_{\rm max} &=& \frac{2\sqrt{2\pi}\Gamma\left(\frac{3}{4}\right)}{3^{3/4}
\Gamma\left(\frac{1}{4}\right)\cosh^{\frac{1}{2}}\eta} \left[1 - \frac{7}{2}
a^4 r_0^4\cosh^2\eta + \cdots\right]\nn
&=& \frac{0.74333}{\cosh^{\frac{1}{2}}\eta}\left[1 - \frac{7}{2}
a^4 r_0^4\cosh^2\eta + \cdots\right] 
\eea
By using \eqref{gaugegravity} we can rewrite $\ell_{\rm max}$ in \eqref{lmax}
in terms of the gauge theory parameters as,
\be\label{lmaxgauge}
\ell_{\rm max} = 0.74333 (1-v^2)^{\frac{1}{4}}\left[1 - \frac{7}{2}
\frac{\pi^4 \hat \lambda T^4 \theta^2}{1-v^2} + \cdots\right]
\ee
where we have used $\cosh\eta = \gamma = 1/\sqrt{1-v^2}$, with $v$ being the
velocity of the dipole. In \eqref{lmaxgauge} the term outside the square
bracket is the commutative result (when we put $\theta=0$) and represents 
the usual $J/\Psi$ suppression of the high velocity quark antiquark produced 
in the 
QGP in the heavy ion collision observed in RHIC 
\cite{Liu:2006nn,Digal:2001iu}. However,
we note that the noncommutativity reduces this result due to the second term
in the square bracket in \eqref{lmaxgauge}. 
The quantity $L_{\rm max} = \ell_{\rm max}/(\pi T)$ can be thought of
as the screening length of the dipole since this is the maximum value of $L$
beyond which we have two dissociated quark and antiquark or two disjoint
world-sheet for which $E(\ell)=0$. As the screening length gets smaller less
and less dipoles will be created and there will be more suppression of
quark-antiquark bound states like $J/\Psi$. Noncommutativity makes the
interaction between the quark and antiquark weaker due to nonlocality and 
that is the
reason it makes the screening length shorter. Note that the velocity of
the dipole has an opposite effect in the correction term due to
noncommutativity, i.e., higher the velocity, lower would be the correction
term due to noncommutativity. Also the correction term is more pronounced at
higher temperature. We would also like to remark that noncommutativity gives a
range for the temperature. We mentioned that the above result
\eqref{lmaxgauge} is valid when $a^4r_0^4 \cosh^2\eta\ll 1$ which in turn 
gives a range for the temperature as,
\be\label{templ}
T \ll \left(\frac{1}{\pi^4 \hat \lambda (1-v^2) \theta^2}\right)^{\frac{1}{4}}
\ee
When the temperature is above this value the expansion in \eqref{lmax} 
will break
down and the screening length will no longer be given by \eqref{lmaxgauge}.
In that case the screening length has to be computed in the opposite limit
where $a^4r_0^4 \cosh^2\eta\gg 1$. However, in this limit we haven't been 
able to write a closed form analytic expression for the screening length. 

\section{Jet quenching parameter}

In section 2, we have discussed the case (a) where the rapidity 
$\eta$ 
remains finite and $\sqrt{\cosh\eta} < \Lambda$. So, the velocity of the
background is in the range $0<v<1$ and the Wilson loop is time-like. Now we 
will
discuss the case (b) where $\sqrt{\cosh\eta} > \Lambda$. In order to 
calculate the
jet quenching parameter we take $\eta \to \infty$ or $v \to 1$, so that the
Wilson loop is light-like and then take $\Lambda \to \infty$. The jet
quenching parameter in the NCYM theory has been calculated before 
\cite{Chakraborty:2011gn} in a different approach, namely, we
calculated it directly in the light-cone coordinates. Here we compute it, 
primarily for completeness, from the general velocity dependent string world
sheet action and then taking the limit $v \to 1$.  
As before we compute the expectation value of the Wilson loop by extremizing
the action. The jet quenching parameter, a measure of
radiative parton energy loss in a quark-gluon plasma medium, is related 
to a particular light-like Wilson loop and thus we obtain its value using the
holographic gauge/gravity duality. Since this has already been calculated in 
\cite{Chakraborty:2011gn}, using a different approach, we will be brief here. 
Note that as
$\cosh^2\eta$ is now greater than $\Lambda^4$, where $\Lambda$ is the upper
limit of $y$, the factor $(y^4 - \cosh^2\eta)$ appearing in the action
\eqref{ngtwo}, \eqref{lagrangian} is negative and the action becomes
imaginary. So, we rewrite the action 
\eqref{ngtwo} as,
\be\label{ngtwoj}
S = \frac{i{\cal T} r_0}{\pi\alpha'}\int_0^{\ell/2} d\sigma {\cal L} =
i{\cal T} T \sqrt{\hat \lambda} \int_0^{\ell/2} d\sigma {\cal L}
\ee
where
\be\label{lagrangianj}
{\cal L} = \sqrt{\left(\cosh^2\eta - y^4\right)\left(\frac{1}{1+a^4r_0^4y^4} +
\frac{y'^2}{y^4-1}\right)}
\ee
As before since the Lagrangian density does not explicitly depend on $\sigma$,
the corresponding Hamiltonian is conserved. Therefore, we have
\be\label{hamiltonian}
{\cal H} = {\cal L} - y'\frac{\partial {\cal L}}{\partial y'} = {\rm const.}
\quad \Rightarrow \frac{\cosh^2\eta - y^4}{(1+a^4r_0^4y^4)
\sqrt{(\cosh^2\eta - y^4)\left(\frac{1}{1+a^4r_0^4y^4} +
    \frac{y'^2}{y^4-1}\right)}} = q_0
\ee      
where we have denoted the constant as $q_0$. \eqref{hamiltonian}
can be solved for $y'$ as 
\be\label{yprimesoln}
y' =
\frac{\sqrt{1+a^4r_0^4q_0^2}\sqrt{(y^4-1)(y_m^4-y^4)}}{q_0(1+a^4r_0^4y^4)}
\ee
where
\be\label{ym}
y_m^4 = \frac{\cosh^2\eta - q_0^2}{1+a^4r_0^4q_0^2}
\ee
On integration, \eqref{yprimesoln} gives us,
\be\label{ellsoln}
\ell = 2\int_0^{\ell/2} d\sigma = \frac{2q_0}{\sqrt{1+a^4r_0^4   
q_0^2}} \int_1^{\Lambda} \frac{1+a^4r_0^4y^4}{\sqrt{(y^4-1)(y_m^4-y^4)}} dy
\ee
Substituting the value of $y'$ from \eqref{yprimesoln} into the action 
\eqref{lagrangianj}, we simplify its form as,
\be\label{sl}
S(\ell) = \frac{i{\cal T}T\sqrt{\hat \lambda}}{\sqrt{1+a^4r_0^4q_0^2}}
\int_1^{\Lambda} \frac{\cosh^2\eta - y^4}{\sqrt{(y^4-1)(y_m^4-y^4)}} dy
\ee
Now in the above since $\ell$ is related to the dipole length as $L =
\ell/(\pi T)$, it is very small compared to the other length scale in the
theory and so, from \eqref{ellsoln} it is clear that $q_0$ is also a small
parameter. In this approximation $q_0$ can be obtained from \eqref{ellsoln} as
\be\label{q0soln}
q_0 = \frac{\ell \cosh\eta}{2}\left[\int_1^{\Lambda}
  \frac{1+a^4r_0^4y^4}{\sqrt{y^4-1}} dy\right]^{-1}
\ee    
In this limit $S(\ell)$ in \eqref{sl} can be expanded as,
\be\label{expansion}
S(\ell) = S^{(0)} + q_0^2 S^{(1)} + {\cal O}(q_0^4)
\ee
where
\bea\label{terms}
S^{(0)} &=& i{\cal T} T \sqrt{\hat \lambda} \int_1^{\Lambda} \frac{
\sqrt{\cosh^2\eta - y^4}}{\sqrt{y^4-1}} dy\nn
q_0^2 S^{(1)} &=& \frac{i{\cal T} T \sqrt{\hat \lambda}}{2} q_0^2 
\int_1^{\Lambda} \frac{1+a^4r_0^4y^4}{
\sqrt{\cosh^2\eta - y^4}{\sqrt{y^4-1}}} dy
\eea
It can be shown \cite{Liu:2006ug,Liu:2006he} that as $q_0 \to 0$, $S^{(0)}$ 
above is equal to $S_0$, the
self-energy of the dissociated quark and antiquark or area of the two disjoint
world-sheet. So, subtracting the self-energy we obtain the action as
\be\label{action1}
S-S_0 = q_0^2 S^{(1)} = \frac{i{\cal T}T\sqrt{\hat\lambda}}{4} \ell^2 \cosh\eta
\left[\int_1^{\Lambda} \frac{1+a^4r_0^4y^4}{\sqrt{y^4-1}}dy\right]^{-1}
\ee
where we have used \eqref{q0soln} and have taken $\eta \to \infty$. Now ${\cal
  T} \cosh\eta$ in \eqref{action1} can be identified as $L^-/\sqrt{2}$, where
$L^-$ is the length of the Wilson loop in the light-like direction. Let us use
the standard relation (for example see, \cite{Kovner:2003zj})
\be\label{standard}
\langle W({\cal C})\rangle = e^{2i(S({\cal C}) - S_0)} \approx
  e^{-\frac{1}{4\sqrt{2}} \hat q_{\rm NCYM} L^- L^2}
\ee
where the factor 2 in the exponent in the second expression is due to the fact
that we are dealing with adjoint Wilson loop.
Here $L\ll 1$ and $\hat q_{\rm NCYM}$ is the jet quenching parameter of the
NCYM theory. Now we can extract its value from \eqref{action1} as,
\be\label{jetquen}
\hat q_{\rm NCYM} = \pi^2 {\sqrt{\hat \lambda}} T^3 \left[\int_1^{\Lambda} 
\frac{1+a^4r_0^4y^4}{\sqrt{y^4-1}}dy\right]^{-1}
\ee
We would like to point out that the above expression of the jet quenching
parameter is actually a formal expression since by taking $\Lambda \to
\infty$, the integral in the square bracket in the jet quenching expression 
diverges. The reason for this divergence, as we mentioned, is that for the
noncommutative case the gauge theory does not live at $\Lambda = \infty$, but
at some finite $\Lambda$ \cite{Maldacena:1999mh,Dhar:2000nj} (see also
\cite{Alishahiha:1999ci,Bigatti:1999iz,Haque:2009hz}). So, the above 
integral needs to be regularized. We have
given the details of the regularization in \cite{Chakraborty:2011gn} and 
here we just give the
results, namely, the regularized integral has the value,
\be\label{regint}
\int_1^{\infty} \frac{1+a^4r_0^4y^4}{\sqrt{y^4-1}} dy =
\left(1+\frac{a^4r_0^4}{3}\right) a_3, \qquad {\rm with}, \quad a_3 = \frac{
\sqrt{\pi} \Gamma\left(\frac{5}{4}\right)}{\Gamma\left(\frac{3}{4}\right)} 
\ee 
Substituting \eqref{regint} in \eqref{jetquen} and expressing $a^4r_0^4$ in
terms of the gauge theory parameters from \eqref{gaugegravity} we obtain,
\be\label{jetquen1}
\hat q_{\rm NCYM} = \frac{\pi^{\frac{3}{2}} \Gamma\left(\frac{3}{4}\right)}
{\Gamma\left(\frac{5}{4}\right)}\sqrt{\hat \lambda} T^3 \left(1+\frac{\pi^4
T^4 \hat \lambda \theta^2}{3}\right)^{-1}
\ee 
So, for small noncommutativity, $\theta \ll 1$, the jet quenching parameter is
given as,
\be\label{smalljetquen}
\hat q_{\rm NCYM} = \frac{\pi^{\frac{3}{2}} \Gamma\left(\frac{3}{4}\right)}
{\Gamma\left(\frac{5}{4}\right)}\sqrt{\hat \lambda} T^3 \left(1-\frac{\pi^4
T^4 \hat \lambda \theta^2}{3} + {\cal O}(\theta^4)\right)
\ee 
Whereas, for large noncommutativity, $\theta \gg 1$, the jet quenching
parameter takes the form,
\be\label{largejetquen}
\hat q_{\rm NCYM} = \frac{3 \Gamma\left(\frac{3}{4}\right)}
{\pi^{\frac{5}{2}}\Gamma\left(\frac{5}{4}\right)}\frac{1}{\sqrt{\hat \lambda} 
T\theta^2} \left(1-\frac{3}{\pi^4
T^4 \hat \lambda \theta^2} + {\cal O}(\frac{1}{\theta^4})\right)
\ee
We note from \eqref{smalljetquen} that for small noncommutativity, when
$\theta \to 0$, we recover the value of the jet quenching parameter of the 
ordinary Yang-Mills plasma obtained in \cite{Liu:2006ug} as expected. 
In this case the
NCYM 't Hooft coupling $\hat \lambda$ reduces to ordinary 't Hooft coupling.
But in the presence of noncommutativity, the value of the jet quenching gets
reduced from its commutative value and the reduction gets enhanced with
temperature as $T^7$. The reduction in the jet quenching for the
noncommutative case can be intuitively understood as follows. The
noncommutativity introduces a non-locality in space due to space uncertainty
and there is no point-like ineraction among the partons and therefore, the
parton energy loss would be less. Also since for small noncommutativity we
have $a^4r_0^4 \ll 1$, this gives a range for the temperature due to
noncommutativity as,
\be\label{tempjqsmall}
T \ll \left(\frac{1}{\pi^4 \hat \lambda \theta^2}\right)^{\frac{1}{4}}
\ee
when the temperature is above this value, the jet quenching expression will no
longer be given by \eqref{smalljetquen}. In that case we have to use the
expression \eqref{largejetquen} which is valid when the temperature is 
given by the limit
\be\label{tempjqlarge}
T \gg \left(\frac{1}{\pi^4 \hat \lambda \theta^2}\right)^{\frac{1}{4}}
\ee
In this case the jet quenching varies inversely with temperature. 

In \cite{Chakraborty:2011gn} the jet quenching for the NCYM theory 
has been calculated by
evaluating the expectation value of the Wilson loop directly in the light-cone
frame. But here we calculate the same for finite value of the rapidity
$\eta$ and then take the limit $\eta \to \infty$ or the velocity of the
background $v \to 1$. Obviously, both methods give the same results.  
 
\section{Conclusion}

To summarize, in this paper we have computed the expectation value of both the
time-like and the light-like Wilson loop of the NCYM theory in (3+1)
dimensions using the gauge/gravity duality and the Maldacena prescription.
The gravity dual background for the NCYM theory is given by a particular
decoupling limit of (D1, D3) brane bound state system of type IIB string 
theory. The
noncommutative directions were taken as $x^2$ and $x^3$, two of the
world-volume directions of the D3-branes. We introduced a fundamental string
in this background as probe whose end points or the dipole consisting of a 
quark and an antiquark lie along one of the
noncommutative directions, namely, $x^2$-direction. The background or the QGP 
medium is assumed to move along $x^1$-direction
with a finite velocity $0<v<1$. This particular configuration is taken for
simplicity. The expectation value of the Wilson loop is
calculated by extremizing the world-sheet area of the F-string in the
background whose boundary is the loop in question. 

Initially we took the
velocity to be less than 1, the case in which the action is real and the
Wilson loop is time-like. In this case we computed the quark-antiquark
separation length $\ell$ as a function of constant of motion $q$ and using
this we also computed the quark-antiquark potential $E$ as a function of 
$\ell$. We found that unlike in commutative case, $\ell(q)$ diverges as we
take the boundary theory to be living at $\Lambda = \infty$ and so we needed to
regularize an integral to make the result finite. After this regularization we
found that the constant of motion $q$ can not take arbitrarily large values as 
in the commutative case, but it must have a cut-off. However, we could not get
a closed form analytic expression for $\ell(q)$ in general and therefore,
obtained it numerically and plotted the function $\ell(q)$ vs $q$ with fixed
value of the noncommutativity parameter and different values of rapidity and
also with fixed value of rapidity and different values of noncommutativity
parameter. We
discussed the various cases and pointed out the differences with the
commutative results. Similarly, the quark-antiquark potential $E(\ell)$ was
also found to be divergent even after subtracting the self-energies of the
quark and the antiquark unlike in the commutative case and a regularization was
needed. Again in this case we could not get an analytic expression for
$E(\ell)$ and we have plotted the function $E(\ell)$ vs $\ell$ with fixed
value of noncommutativity parameter and varying rapidity as well as fixed
value of rapidity and varying noncomutativity parameter. Here also we
discussed the results and compared them with the commutative results. Then we
gave an analytic expression for the screening length in a special limit,
namely, when the rapidity is large and the noncommutativity parameter is small
with the product remaining small and discussed our results.  

Finally, we discussed the case when the rapidity goes to infinity or the
velocity of the medium approaches 1. In this case the action becomes imaginary
and the Wilson loop becomes light-like. Using this Wilson loop we 
obtained the expression of the jet quenching parameter in the NCYM theory
which was obtained before \cite{Chakraborty:2011gn} using different approach. 
We obtained the expressions 
for the jet quenching parameter for small and large noncommutativity. For
small noncommutativity the jet quenching got reduced from its commutative
value by an amount proportional to $\hat \lambda^{3/2} T^7 \theta^2$, whereas
for large noncommutativity the leading order term was found to be proportional 
to $1/(\sqrt{\hat \lambda} T \theta^2)$. 
   
\section*{Acknowledgements}

We would like to thank Munshi G Mustafa for several discussions. We would also
like to thank the anonymous referee for raising some points, the
clarifications of which, we hope, has improved the manuscript.

\end{document}